\documentclass[usenatbib,structabstract,final]{aa}
\usepackage{amsmath,amsfonts,amssymb,amsxtra} % AMSTEX unterstuetzen
\usepackage{hyperref}
\usepackage{txfonts}
\usepackage{xspace}
\usepackage{graphicx}
\usepackage{natbib}
\usepackage{color}

\newcommand{\logg}{\log g }
\newcommand{\Msun}{\,$\rm{M}_\odot$\xspace}
\newcommand{\vsini}{\varv \sin i }
\newcommand{\vzams}{\varv_{\mathrm{ZAMS}} }

\newcommand{\kms}{$\rm{km/s}$\xspace}
\newcommand{\Teff}{\rm{T}_{\rm{eff}}}

\def\mso{\,{M}_{\odot}}

\defcitealias{Brott10_gridpaper}{Paper~I}

\begin{document}

    \title{Rotating massive main-sequence stars II: \\
     Simulating a population of LMC early B-type stars as a test of rotational mixing}
    \author{ I. Brott \inst{1,2} \and C. J. Evans \inst{3} \and  I. Hunter\inst{4} \and A. de Koter \inst{5,1} \and 
     N. Langer \inst{6,1} \and P. L. Dufton\inst{4} \and M. Cantiello \inst{6} \and 
     C. Trundle \inst{4} \and \newline D. J. Lennon  \inst{7} \and S.E. de Mink \inst{6-8} \and  S.-C. Yoon \inst{6} \and P. Anders \inst{1}}

    \institute {Astronomical Institute, Utrecht University, Princetonplein 5, 3584 CC, Utrecht, The Netherlands 
    \and University of Vienna, Department of Astronomy, T\"urkenschanzstr. 17, A-1180 Vienna, Austria
    \and UK Astronomy Technology Centre, Royal Observatory Edinburgh, Blackford Hill, Edinburgh, EH9 3HJ, UK
    \and Astrophysics Research Centre, School of Mathematics \& Physics, The Queens University of Belfast, Belfast, BT7 1NN, Northern Ireland, UK
    \and Astronomical Institute Anton Pannekoek, University of Amsterdam, Kruislaan 403, 1098 SJ, Amsterdam, The Netherlands
    \and Argelander-Institut f\"ur Astronomie der Universit\"at Bonn, Auf dem H\"ugel 71, 53121 Bonn, Germany 
    \and Space Telescope Science Institute, 3700 San Martin Drive, Baltimore, MD 21218, USA
    \and Hubble Fellow }

\date{Received:  / Accepted: }

  \abstract
    {Rotational mixing in massive stars is a widely applied concept, 
    with far-reaching consequences for stellar evolution,
    nucleosynthesis, and stellar explosions.} 
    {Nitrogen surface abundances for a large and homogeneous sample 
    of massive B-type stars in the Large Magellanic Cloud (LMC) have
    recently been obtained by the ESO VLT-FLAMES Survey of Massive Stars. This sample 
    is the first to cover a broad range of projected stellar rotational velocities,  
    with a large enough sample of 
    high quality data to allow for a statistically significant analysis.
    Here we use the sample to provide the first
    rigorous and quantitative test of the theory of rotational mixing in
    massive stars.}  
    {We calculated a grid of stellar evolution models, using the VLT-FLAMES sample
    to calibrate some of the uncertain mixing processes.
    We developed a new population-synthesis code, which 
    uses this grid to simulate a large population of stars with masses, 
    ages, and rotational velocity distributions consistent with those 
    from the VLT-FLAMES sample. The synthesized population is then filtered by the 
    selection effects in the observed sample, to enable a direct comparison between the 
    empirical results and theoretical predictions.}
    {Our simulations reproduce 
    the fraction of stars without significant nitrogen enrichment. However, the predicted
    number of rapid rotators with enhanced nitrogen is about twice as large as found observationally.   
    Furthermore, two groups of stars (one consisting of slowly rotating, nitrogen-enriched objects and
    another consisting of rapidly rotating un-enriched objects) cannot be 
     reproduced by our single-star population synthesis.}  
    {Physical processes in addition to rotational mixing appear to be required to understand
    the population of massive main-sequence stars from the VLT-FLAMES sample.
    We discuss the possible role of binary stars and magnetic fields in the interpretation of our results.
    We find that the population of slowly rotating nitrogen-enriched stars is unlikely to be 
    produced via mass transfer and subsequent tidal spin-down in close binary systems. 
    A conclusive assessment of the role of rotational mixing in massive stars requires a quantitative
   analysis that also accounts for the effects of binarity and magnetic fields.}


\titlerunning{Simulating a Population of LMC early B-type Stars as a Test of Rotational Mixing}

\maketitle

%: ------------------- INTRODUCTION -----------------------------------
\section{Introduction}

Observations of chemically enriched main-sequence stars
 \citep[e.g.][]{Gies92, Herrero93, Vrancken00,Przybilla10}
 have fostered the idea that rotationally induced mixing
may bring fusion products from the core of massive stars to their 
surface during core hydrogen burning.  
While the potential consequences of rotational mixing for the
evolution of massive stars have been shown to be dramatic \citep{Maeder00,Yoon05},
direct tests of the models have proven to be difficult.
A major obstacle of such tests is that observations have typically been restricted to
stars that have low projected rotation rates, to minimize any rotational broadening of the stellar absorption lines.
However, the often small sample size has allowed speculation that 
nitrogen-enriched stars might be rapid rotators viewed nearly pole-on.
Since the range of observed enrichments was consistent with evolutionary predictions of
rapidly rotating stars, this was considered as confirming rotational mixing in massive
stars \citep{Heger00b,Meynet00}.

It was only in the framework of the ESO VLT-FLAMES Survey of Massive Stars \citep{Evans05_gal}
that nitrogen abundances were determined for massive main-sequence stars with 
a wide range of projected rotation rates \citep{Hunter07_chem_Bstars,Trundle07}.
Further analysis of the early B-type stars in the Large Magellanic Cloud (LMC) 
from this survey revealed the existence
of a significant population of nitrogen enhanced stars with {\em intrinsically} slow
rotation \citep{Hunter08_letter}. The findings of \citet{Morel06}
argue for a similar population in our Galaxy. The origin of the nitrogen enhancement
in these stars is not understood, but, since they are slow rotators, it seems doubtful
that rotational mixing can explain it. It is therefore conceivable that 
samples of enriched stars, with apparently slow rotation velocities (which substantiated the most
convincing direct evidence for rotational mixing in massive main-sequence stars over
the past decades), may instead be related to completely different physical processes
-- e.g. binarity \citep{Langer08_iaus} or magnetic fields \citep{Morel08}.
In this context, it is worth noting that a wealth of atmospheric
and wind properties of massive main-sequence stars are yet to be understood,
such as the winds of massive stars with luminosities below $\sim$$10^{5.2}\,$L$_{\odot}$ 
\citep[e.g][]{bou03,Martins04,mar05,Mokiem07,Puls08}, intermittent discrete absorption components 
in UV absorption lines \citep{Prinja88, Kaper97},
micro- and macro-turbulence \citep{Cantiello09,Aerts09}, 
and non-thermal X-ray emission \citep{Babel97,ud-Doula02}.

Putting the problem of the enriched slow rotators aside, 
both \citet{Hunter08_letter} and \citet{Maeder09_coast} suggested that
the majority of the newly discovered population of rapidly rotating,
nitrogen enhanced main-sequence stars in the VLT-FLAMES survey may 
agree with the predictions of rotational mixing.
However, their results remain ambiguous, as the existence of a
population of rapidly rotating, nitrogen enriched, apparently single stars
is also a prime prediction of close binary evolution models accounting
for rotation \citep{Petrovic05, Langer08_iaus}, which is independent
of the mechanism of rotational mixing.

In this context, a quantitative effort to reproduce the physical
properties of the LMC B-type sample from the VLT-FLAMES survey (see
Sec.~\ref{sec:obs_sample}) using stellar evolution models for single stars which
allow for the effects of rotational mixing, is a logical next step.
A future step will be to also account for binary evolution.
To achieve the present aim, we use a dense grid of stellar evolution models for
main-sequence stars \citep[Paper~I]{Brott10_gridpaper}, for which
the initial abundances, convective core overshooting, and rotational
mixing were calibrated using the results from the VLT-FLAMES survey. These
models are then incorporated in a newly developed population-synthesis
code (Sec.~\ref{sec:popsyn} ), which we employ in an attempt to reproduce
the properties of the LMC sample of B-type stars from the VLT-FLAMES survey.
The observed sample is briefly discussed in Sec.~\ref{sec:obs_sample}.
Our results are described in Sec.~\ref{sec:results}. We draw our
conclusions in Sec.~\ref{sec:discussion}.

\section{Observational sample}
\label{sec:obs_sample}
To investigate the predictions of rotational mixing, in this paper we
focus on population synthesis of the 107 main-sequence B-type stars in
the LMC discussed by \citet{Hunter08_letter,Hunter09_nitrogen}. These were taken from the
VLT-FLAMES observations in the two LMC fields -- N11 and NGC\,2004
\citep{Evans06_lmc_smc}. N11 is the second largest H\,{\sc ii} region
in the LMC, with multiple clumps of star formation in which the VLT-FLAMES
observations sample a range of stellar ages and spatial structures.
NGC\,2004 is an older, fairly condensed cluster, in which the targets
were mostly in the outer region (the core is too densely populated for
the FLAMES-Medusa fibres) and out into the surrounding field
population. A total of 225 O- and early B-type stars were observed with
VLT-FLAMES in the two LMC pointings.

To constrain our synthesis models we need to understand the selection
effects which influenced the original observed sample and the
subsample considered by \citet{Hunter08_letter}. The most significant
factor in the observed sample was the faint cut-off ($V \le 15.5^{\rm
  m}$, to ensure sufficient signal-to-noise in the final spectra),
combined with functional issues such as crowding of potential targets.

In the NGC\,2004 field we note that, contrary to the statement by
\citet{Evans06_lmc_smc}, some of the Be-type stars identified by
\citet{Keller99} were actually hardwired out of the input catalog
used for target selection.  This could potentially bias the final
distribution of Be- relative to B-type stars in this field. 
 However, given the magnitude (and color) constraints on targets,
only 18 such stars would have been included in our catalog of potential targets.
\citet{Hunter08_vrot} investigated the potential for similar selection effects in NGC\,330 (in the SMC)
using the Fibre Positioner Observation Support Software (FPOSS) with the inclusion of previously excluded stars. We have adopted a similar method for
NGC\,2004, running FPOSS ten times to compare the resulting configurations. Of the previously excluded
stars from \citet{Keller99}, six stars were included in the resulting fibre configuration in two of the
test runs, seven stars were included in six of the tests, and eight stars were included in the remaining
two tests.  Of course, by including some of these previously excluded stars, a corresponding number of
the stars actually observed would not have been included, some of which might well be Be-type stars (18
of the 116 stars observed with the Medusa fibres in NGC\,2004 are Be-type, i.e. 15\%). 
There may be a weak selection effect relating
to the observed sample of Be-type stars, but we believe that this does
not unduly bias the overall sample in NGC\,2004 -- other effects such
as the spatial distribution of targets and the magnitude/color cuts
would have been more significant.  In short, the survey provided a
relatively unbiased sample of the bright ($V \le 15.5^{\rm m}$)
main-sequence stars in the observed regions.

\subsection{Stellar parameters}
Physical properties of the LMC B-type stars were determined in a
series of papers. The stellar parameters used in our LMC population
synthesis (N11 and NGC\,2004) are those from
\citet{Hunter08_vrot}, which incorporated results for narrow-lined
B-type stars from \citet{Hunter07_chem_Bstars} and \citet{Trundle07},
\citep[and for O-type stars from][]{Mokiem06}.

Physical parameters and abundances for the B-type stars were
obtained using the grid of non-LTE model
atmospheres described by \citet{Dufton05}, calculated using the {\sc
  tlusty} code \citep{Hubeny95}.  Effective temperatures were
estimated from the silicon or helium spectrum, while in those stars
where adequate spectral diagnostics were not available, temperatures
were adopted on the basis of their spectral classifications using the
calibration from \citet{Trundle07}.  Surface gravities ($\logg$) were
deduced from the hydrogen line profiles. Projected rotational
velocities ($\vsini$) were determined using the profile-fitting
methods of \citet{Dufton06_vrot}, i.e. using the helium lines
(primarily He\,I at $\lambda$4026\,\AA) in the majority of stars and
using metal lines (such as Mg\,{\sc ii}, Si\,{\sc iii}) at low
$\vsini$.  Surface abundances of C, N, O, Mg, and Si were
also determined where possible \citep{Trundle07,Hunter07_chem_Bstars,
Hunter08_letter,Hunter08_vrot,Hunter09_nitrogen}.

Results for 107 main-sequence B-type stars were given by
\citet{Hunter09_nitrogen}.  From the original sample of 225 O- and early
B-type stars: 41 were O-type; 1 was a W-R star; 24 were Be-type (for
which N abundances were not determined); 23 additional B-type stars
were not analyzed due to doubled-lined binarity, 'shell-like'
emission, and low-quality spectra; 29 stars have $\logg$\,$<$\,3.2.
Note however that the observations of the VLT-Flames survey were
not optimized for detection of binaries, so some almost certainly
remain undetected.  In Appendix \ref{App:A} we provide mass
and velocity distributions of the observational sample.
We refer to Sec.~\ref{sec:selectioneffects} for a description of how the selection effects
are applied to the population synthesis model.

%: ----------- POPULATION SYNTHESIS ------------------------
\section{Stellar models and population synthesis}
\label{sec:popsyn}
We have developed the population synthesis code {\sc STARMAKER} to study the statistical properties 
of early-type stars for a specified initial rotational velocity,
initial mass function (IMF), and given star formation history (SFH). 
This method can be applied to quantitatively constrain the effects of rotation (and mass loss) on
stellar evolution if the dataset to which its results are compared is sufficiently large to provide good sampling of the (random) orientation of 
rotation axes, and of the velocity and age distribution. 
On a  given grid of stellar evolution models, {\sc STARMAKER} interpolates model stars of a specified initial mass, age and initial rotational velocity within the grid. 
In addition to the main stellar parameters (effective temperature, luminosity, radius and surface gravity), 
the following quantities are also obtained: mass-loss rate, current rotational velocity, 
current stellar mass, and surface abundances (He, B, C, N, O, Ne, Na, Si).  
For the simulation presented in this paper, we use the LMC stellar evolution grid  presented in 
\citetalias{Brott10_gridpaper} as input for our population synthesis models. \\
In Sec.~\ref{sec:popsyn_methodes} we discuss the implementation method applied, and justify the initial distributions in Sec.~\ref{sec:initial_distributions}. 
The selection effects applied to both the observed and modeled population are reviewed in 
Sec.~\ref{sec:selectioneffects}. Note that some aspects of this 
selection effects are linked to the discussion in Sec.~\ref{sec:initial_distributions}. Sec.~\ref{sec:stev} and \ref{sec:rotmix} discuss stellar evolution models and rotational mixing. 

%:------------------- Population Synthesis details ------------------------
\subsection{Population synthesis }
\label{sec:popsyn_methodes}

\begin{figure}[htbp]
\begin{center}
\includegraphics[angle=-90,width=0.5\textwidth]{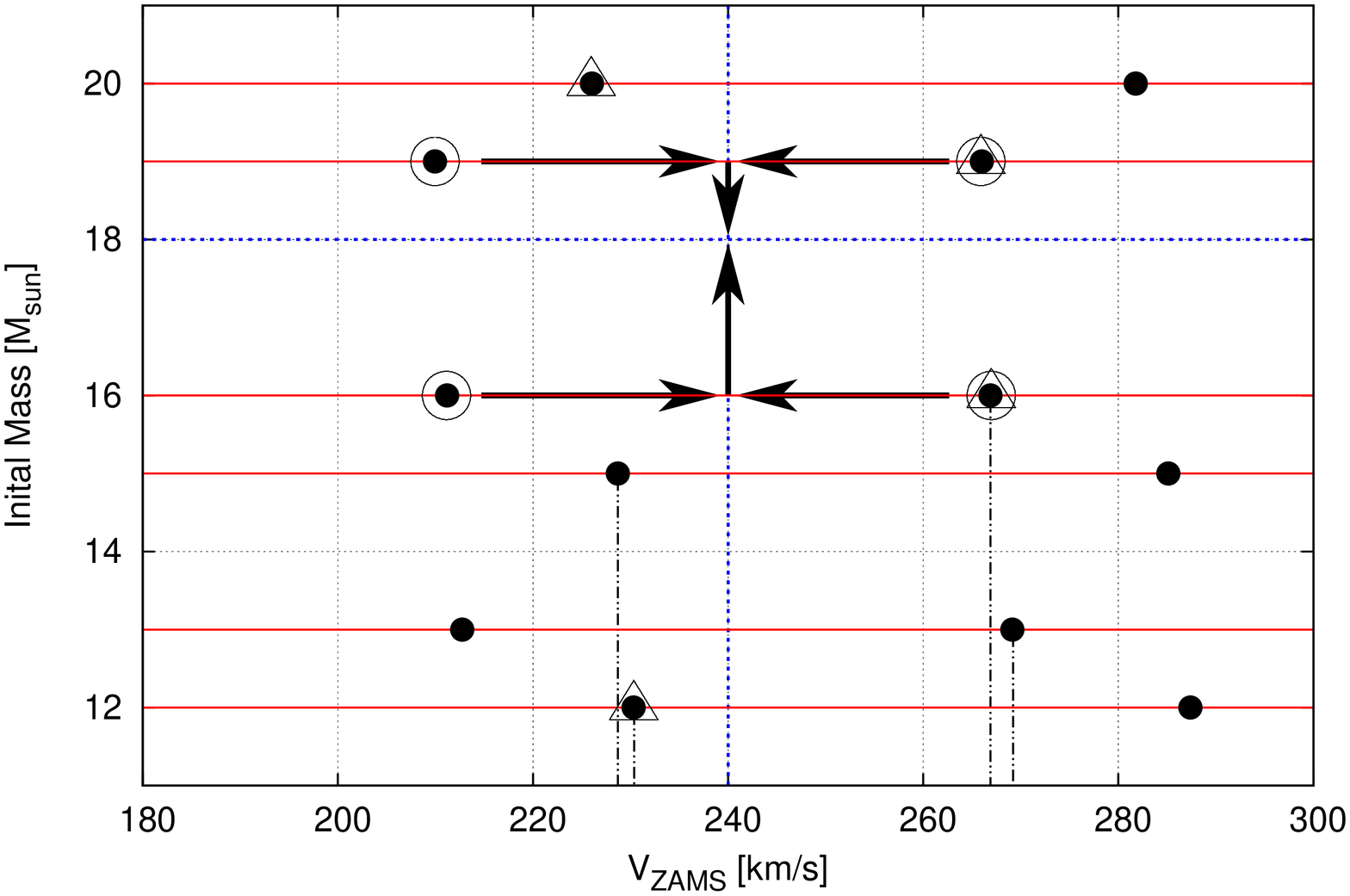}
\includegraphics[angle=-90,width=0.5\textwidth]{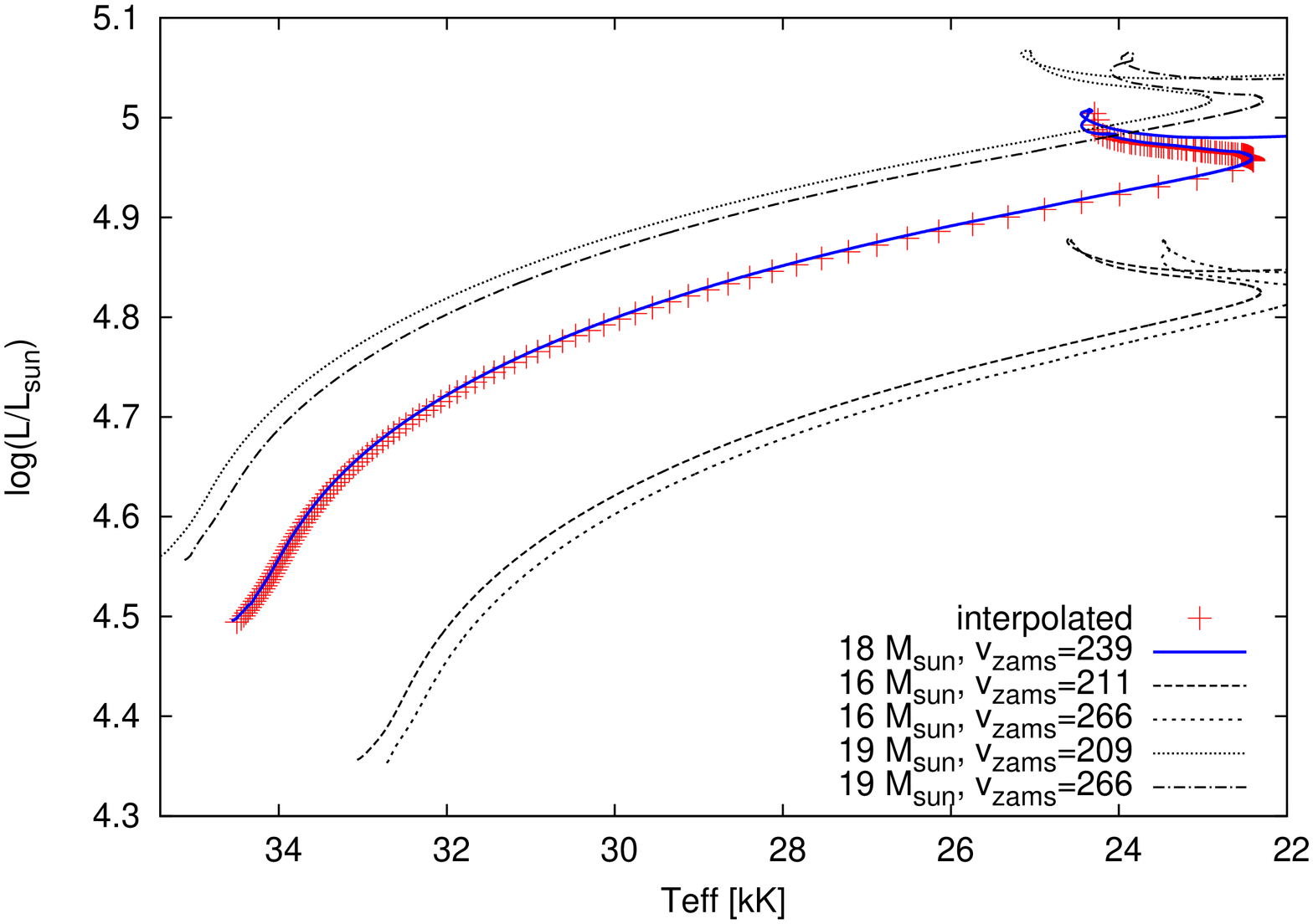} 
\caption{
Top: Illustration of the interpolation procedure of STARMAKER in the
$\vzams$ -- $\rm{m}_{i}$ plane.
Individual model sequences are indicated by filled circles. 
Red lines indicate lines of constant mass.
In this example, the properties of a star with initial mass $M_{\rm i}=18$\Msun and initial  velocity $\vzams=240\,\kms$
are determined through interpolation (the intersection of the blue dashed lines).  
{\sc STARMAKER} chooses the circled models when interpolating. 
Models marked by triangles would be chosen if the nearest neighbor method was used. 
Black arrows indicate the sequence of the interpolation, as described in the text. Dashed lines are intended as guide to the eye to distinguish models of similar velocity.
Bottom: Interpolated track (red crosses), resulting from the example given above. The over plotted full blue line shows the model calculated with our stellar evolution code. The four black lines show the evolutionary tracks from which the model has been interpolated. }
\label{fig:model_selection}
\end{center}
\end{figure}

STARMAKER simulates a prescribed number of stars, with
initial mass ($M_{\rm i}$), initial rotational velocity ($\vzams$), and age ($t$) drawn from the corresponding distribution function (IMF, initial distribution of rotational velocities, and SFH). 
For a given initial triplet ($M_{\rm i}$, $\vzams$, and $t$), four evolutionary sequences from the grid of stellar evolution calculations are selected, 
and then used to interpolate the stellar parameters using a 3D linear interpolation algorithm. 

For a given $M_{\rm i}$ and $\vzams$,
the interpolation routine first determines the two closest masses above and below $M_{\rm i}$ in the model grid. 
Then, for each of the two, it chooses the two models with $\vzams$ closest 
to the required value (Fig.~\ref{fig:model_selection}). 
We found that the algorithm works best when the four selected model sequences form two pairs with equal masses. 
For example, to interpolate an 18\Msun model, two 16\Msun and two 19\Msun models enter the interpolation.  
An interpolated track is show in Fig.~\ref{fig:model_selection} (bottom). Initial masses and initial velocities may be irregularly spaced in the model grid, but for every new mass introduced into the model grid, models of several rotational velocities are required to cover the $\vzams$-parameter space considered in the simulation. 

As indicated by the arrows in Fig.~\ref{fig:model_selection}, we first interpolate models of equal mass
to obtain models with the desired initial velocity, and then we interpolate in mass.
We tested our result by changing the interpolation order, and 
found no significant differences.
Adopting the nearest neighbors for interpolation , i.e the smallest distance $\sqrt{(\Delta M)^{2}+(\Delta \vzams)^{2}}$,
would often result in selecting models of three or four different initial masses  due to the domination of the 
velocity term in the distance equation 
(models marked by a triangle in Fig.~\ref{fig:model_selection}). We found this
could lead to unsatisfactory results.

\begin{figure}[htbp]
\begin{center}
\includegraphics[angle=-90,width=0.5\textwidth]{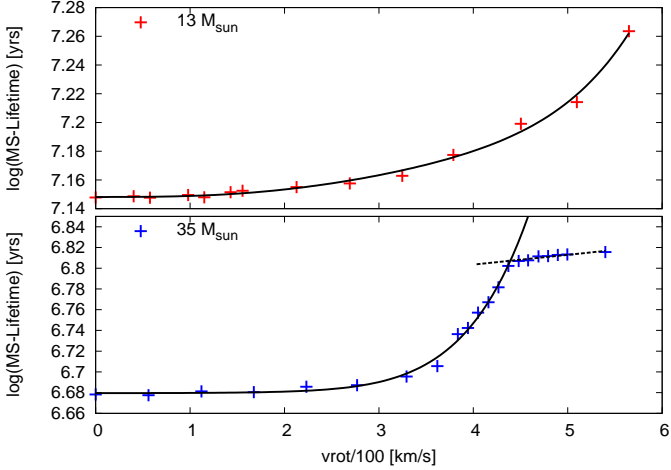}
\caption{The logarithm of the MS lifetime is plotted against the ZAMS rotational velocity for our 13\Msun (top) and 35\Msun(bottom)
sequences. In black we show the polynomial fit to the model data.  
The bottom panel shows also the linear part (dashed line) used to fit the models evolving quasi-chemically homogeneously. }
\label{fig:fitting_age}
\end{center}
\end{figure}

Stars with different initial masses and rotation velocities achieve different main-sequence lifetimes. 
To facilitate the interpolation in time, we use the fractional main-sequence lifetime as the 
interpolation variable. For all models in the grid their MS-lifetimes have been tabulated so that, for any pair ($M_{\rm i}$, $\vzams$), 
the expected MS-lifetime can be interpolated. To ensure a smooth behavior of isochrones in the HR diagram
we fit the MS-lifetime for each initial mass as function
of $\vzams$ with a polynomial, and add a linear fit for masses where quasi-chemically homogeneous evolution 
is achieved for the highest rotational velocities (Fig.~\ref{fig:fitting_age}). 
The increase in the MS-lifetime for higher rotational velocity
is due to rotational mixing of hydrogen into the convective core.
This effect saturates when the rotational mixing timescale becomes shorter than the main sequence timescale.
At this point the star becomes quasi-chemically homogeneous \citep[e.g.][]{Maeder87, Yoon05}, 
implying its interior is well mixed and the star can burn basically all the fuel it contains.

Any observable can now be interpolated using the four model sequences
used for interpolation (Fig.~\ref{fig:model_selection}). As each model sequence is highly resolved in time
(with several thousand time steps for the main sequence evolution), the interpolation error in time is negligible.

%: ------------- Initial Distributions -------------------------------------------
\subsection{Initial distribution functions}
\label{sec:initial_distributions}
In our population synthesis we draw the three initial parameters ($M_{\rm i}$, $\vzams$, and $t$), 
from the initial distribution functions, which are discussed below.  We also adopt an inclination angle for the star, assuming
random orientation in space.  

\subsubsection{Star formation history} \label{sec:sfh} 

\begin{figure}[ht]
\centering
\includegraphics[angle=-90,width=0.5\textwidth]{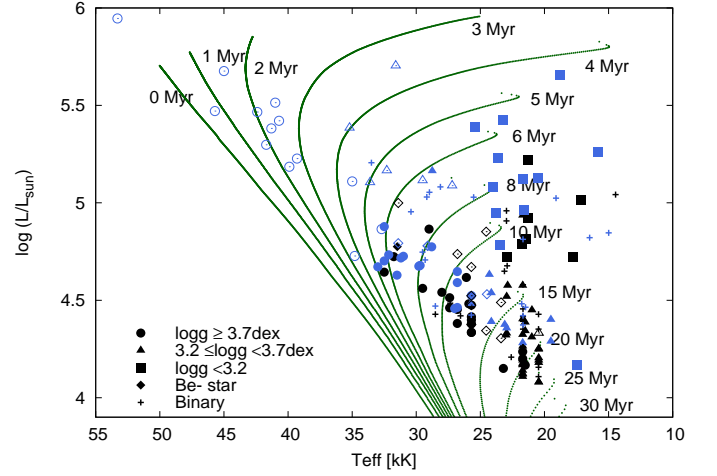}
\caption{\label{fig:isochrones} Isochrones generated from our non-rotating evolutionary models, 
compared to the location of stars of our sample,
in the H-R diagram. Stars for which nitrogen abundances are available 
are shown with filled symbols, while we used empty symbols for those for which this is not the case. 
Black and blue symbols show data from NGC\,2004 and N11, respectively. 
}
\end{figure}

We have used our population synthesis code 
to generate isochrones for non-rotating stars with ages between 0 and 30 Myrs from our stellar evolution grid 
\citepalias[see][]{Brott10_gridpaper}.  In order to 
evaluate the appropriate star formation history for our population synthesis model,
we compare those isochrones with the distribution of the stars of our observational sample in Fig.~\ref{fig:isochrones}.

The N11 region (blue symbols) contains a number of stars with ages
younger than $\sim$5\,Myr. These are primarily the massive O-type
stars, associated with the recent star formation in, e.g., LH\,10.
Both fields contain older populations with ages in the
range of 5 to 25\,Myr, and with no obvious preference for a particular
age (Fig.~\ref{fig:isochrones}). Given that we do not include the O-type
stars in our analysis (see Sec.~\ref{sec:selectioneffects}), we therefore adopt continuous 
star formation as a good approximation of the observed lower-mass population 
in our population synthesis models.
 
%: IMF -----

\begin{figure}[htbp]
   \centering
  \includegraphics[angle=-90,width=0.5\textwidth]{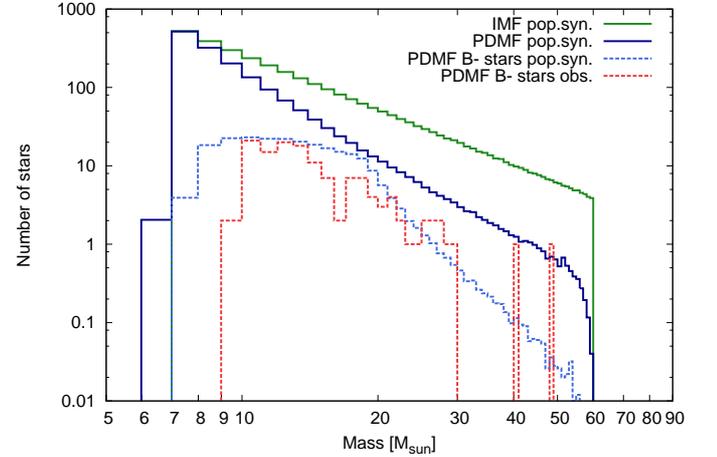}
  \caption{Mass functions from the observed sample and our population synthesis simulation. The input initial mass function (IMF) used in the population synthesis is shown in green. The blue solid line represents the present day mass function (PDMF) of all stars in the simulation. The blue dashed line gives the PDMF of the stars which successfully pass our selected effect filters, in comparison the PDMF of all B-type stars of the observed LMC sample, including those for which no nitrogen measurement is available (red line) .  The synthesized mass functions 
have been scaled by a factor of 1/500 for ease of comparison with the observed distribution.}
 \label{fig:IMF}
\end{figure}
  
\subsubsection{Initial mass function}   \label{sec:imf} 
As input for our population synthesis we adopted a \citet{Salpeter55} IMF. 
The masses are randomly picked from the IMF in the range between 7 and 60\Msun, and the result is plotted in Fig.~\ref{fig:IMF} (green, solid line). 
The  present day mass function (PDMF) of our population synthesis simulation is also plotted (blue, solid line). It contains only main sequence
stars and is calculated assuming continuous star formation. A comparison between the IMF and PDMF shows that mass loss
has populated the 6\Msun-bin and that the (negative) slope of the PDMF has increased. The latter is an evolutionary effect: the most massive stars evolve
more quickly and once they leave the MS they are removed from the population.
At the high mass end the PDMF gets incomplete as mass loss shifts stars to lower mass bins 
and we do not have stars initially more massive than 60\Msun in our simulation. 

To test if our assumptions on the IMF and the SFH are reasonable, 
we construct the PDMF of the stars that would pass all observational selection effects
in our population synthesis simulation
(see Sec.~\ref{sec:selectioneffects}  below).
This distribution corresponds to the light-blue dashed line in Fig.~\ref{fig:IMF}, 
and is compared to the PDMF of the stars in our sample (red line). 
A plateau at lower masses develops, since many low mass stars are too faint to be included in the observational sample. 
At higher masses, the slope steepens further, since the hot O-type stars drop out of the sample (see Sec.~\ref{sec:selectioneffects}). 
This can be directly compared to the observed sample PDMF, shown as a red, dashed line in  Fig.~\ref{fig:IMF}.  
Keeping in mind the low number statistics in the observed sample, it appears that
below 30\,\Msun the two curves have approximately the same shape, thus supporting the assumptions
made for the simulation.

The lowest mass stars in the observed sample are 9\Msun while the lowest mass stars in the simulation are 7\Msun. 
This is confirmed in Fig.~\ref{fig:selectioneffects}, where no stars
are observed at the lowest masses. This could imply that the SFH is limited to $\sim$25\,Myrs,
or --- perhaps more likely --- that the sample selection concentrated on the brighter stars
in the field. This discrepancy between our model and the observational sample might
introduce an error in our analysis. However, the error caused in the population numbers of the different regions of the Hunter-diagram by including stars older than 25\,Myrs
in the simulation is on the order of 1\%. We discuss this in more detail in Sec.~\ref{sec:uncertainties}.

\subsubsection{Initial velocity distribution}  \label{sec:vinit} 
\begin{figure}[htbp]
   \centering
    \includegraphics[angle=-90,width=0.5\textwidth]{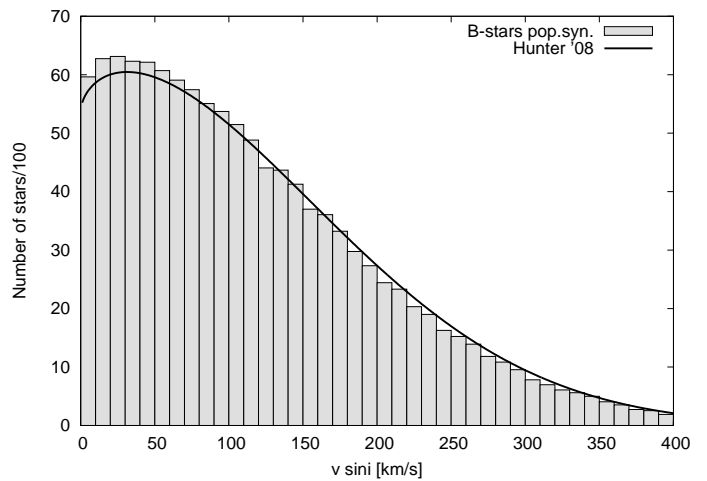}	
   \caption{\label{fig:popsyn_vzams} The projected rotational velocity
     distribution found by \citet{Hunter08_vrot} for non-binary,
     non-supergiant stars of this sample with masses $\leq25$\Msun
     (black line). This is used as our initial rotational velocity
     distribution.  The gray histogram represents the rotational
     velocity distribution of our simulated population, after
     application of all selection effects described in
     Sec.~\ref{sec:selectioneffects}. }
\end{figure}
\citet{Hunter08_vrot} fitted the observed rotational velocity
distribution of the LMC sample with a Gaussian function, excluding
radial velocity variables, supergiants and stars above 25\Msun. 
This mass threshold was used to avoid effects from mass loss 
induced spin-down on the velocity
distribution (see Fig.~\ref{fig:popsyn_vzams}).  As
such, all but the seven lowest mass O-type stars were not included
in determining the velocity distribution. 
Be-stars below the mass threshold are included into the fit to the velocity
distribution.
 As discussed by\citet{Hunter09_erratum}, the width of this function given by
\citet{Hunter08_vrot} is too small and needs to be corrected by a
factor $(\pi/4)^{1/4}$.  Taking this into account, our ZAMS rotational
velocities are selected randomly from the Gaussian function given in
Eq.~\ref{eq:gauss} (using $\mu=100\,\kms\ \rm{and}\
\sigma=141\,\kms$), which has been truncated at zero $\kms$ and
renormalized (see \citealt{Hunter08_vrot}).
 
\begin{equation}
f(\varv) \propto \exp\left[-\frac{1}{2}\left(\frac{\varv-\mu}{\sigma}\right)^{2}\right] .
\label{eq:gauss}
\end{equation}

By using this function for our population synthesis, 
we assume that it is representative of the distribution function of the {\em initial} rotational velocities.
This is justified when the stars rotate close to solid body rotation, 
which is ensured by including magnetic fields in our models. Also,
due to the low mass loss rates for the bulk of the stars in the sample, 
our models show little change of the rotational velocity over the main sequence evolution. 
To test this assumption, we have constructed a histogram of
the present day  projected rotational velocity distribution (PDVD) 
for the B-type stars  from our population synthesis (gray bars in Fig. \ref{fig:popsyn_vzams}). 
Comparing this for the initial  projected rotational velocity distribution (black line)  from \citet{Hunter08_vrot}, 
our simulation predicts only slightly more slow rotators than the input function, 
but also slightly fewer fast rotators around 200\,km/s. 
The overall shape at lower rotational velocities is preserved in our simulated data. 
Overall, the observed B-type star velocity distribution seems to be a good approximation to the  initial velocity distribution of our population synthesis simulation.

%:-------------------- Selection Effects -------------------------------------
\subsection{Selection effects }
\label{sec:selectioneffects}

\begin{figure*}[htbp]
\centering
\includegraphics[angle=-90,width=\textwidth]{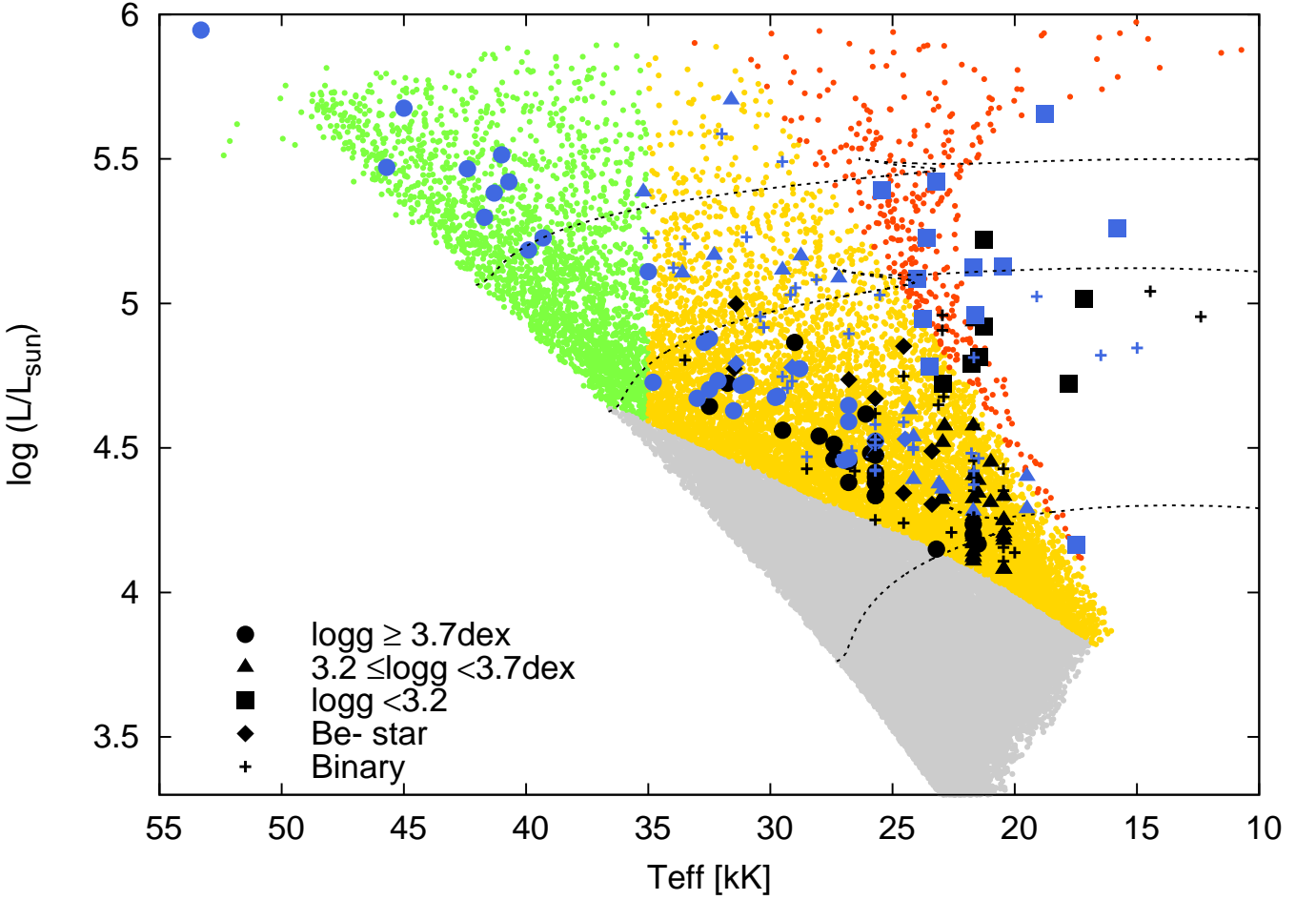}
\caption{
H-R diagram of our population synthesis using the initial distributions discussed in
Sec.~\ref{sec:initial_distributions}. Regions subject to selection effects (Sec.~\ref{sec:selectioneffects}) are highlighted as: gray -- stars too
faint for the observed magnitude cutoff; green -- stars with $\Teff \,>\, 35\,\mathrm{kK}$; red -- stars
with low surface gravities ($\logg\,<3.2$); yellow -- remaining stars.
Results from \citet{Hunter09_nitrogen} are over-plotted (blue: stars in N11; black: NGC\,2004; with their shapes 
indicating their surface gravities).
Radial velocity variables and Be-stars that are indicated by crosses and diamonds, respectively. 
Evolutionary tracks of 10\Msun, 20\Msun and  30\Msun of non-rotating models are shown for orientation.
}
\label{fig:selectioneffects}
\end{figure*}

The VLT-FLAMES survey was designed to contain a largely unbiased sample of O- and B-type stars.
However, owing to both observational and analytical constraints, there are a number of selection
effects present in the final \citet{Hunter08_vrot} sample (see~\ref{sec:obs_sample}), which we attempt to
include in our simulation.
Specific affected regions in the H-R~diagram  are highlighted in Fig.~\ref{fig:selectioneffects}.

\subsubsection{Magnitude limit}
Our LMC sample is magnitude limited, to ensure sufficient signal-to-noise
in each observed spectrum, with about $V = 15.53^{\rm m}$ for the faintest star.
While the limit is not exactly the same
in N11 and NGC2004, the limits in each are not
significantly different and hence it is reasonable to model these together.
In our simulation,
we therefore remove all simulated stars that are fainter than this threshold from our simulation, with
this region marked in gray in Fig.~\ref{fig:selectioneffects}. Magnitudes for the simulated stars
have been calculated using: 
\begin{equation}
\rm{V} =\rm{M}_{\rm{bol}}^{*}+\mu+\rm{R}_{\rm{v}}\cdot E(B\,-\,V) - BC
\end{equation}
\begin{equation}
\rm{M}_{\rm{bol}}^{*}=  -2.5\cdot \log\rm{L_{Model}}+\rm{M}_{\rm{bol},\odot}
\end{equation}
where $\rm{M_{\rm{bol}}^*}$ is the absolute bolometric
magnitude of the simulated star.  Observational parameters are adopted such
that they are consistent with those used by \citet{Hunter08_vrot} for NGC\,2004,
i.e., distance modulus ($\mu$)\,$=$\,18.56$^{\rm m}$, $E(B-V)$\,$=$\,0.09$^{\rm m}$,
and a ratio of total-to-selective extinction $\rm{R}_{\mathrm{V}}$\,$=$\,3.1.
Bolometric corrections (BC) for $\Teff\, <\, 28172\,\mathrm{K}$ have been
adopted from \citet{Balona94}, and for $\Teff\,\ge\, 28172$ from \citet{Vacca96}. 

\subsubsection{Exclusion of O-type stars}
The dominant nitrogen lines
in O-type stars are those from N\,{\sc iii}. However, this ion is
notoriously difficult to model in non local thermodynamic equilibrium
due to the need to include di-electronic recombination channels. To
date, no large studies of nitrogen abundances in O-type stars are
available, although He abundances were studied by \citet{Mokiem06,Mokiem07}.
Due to the lack of nitrogen abundances for stars hotter than 35\,kK we
exclude such hot stars from the simulation (green group in
Fig.~\ref{fig:selectioneffects}), corresponding to spectral types
earlier than O9.

\subsubsection{Exclusion of low gravity objects}
The bulk of the
stars have masses in the range of 10 to 20\Msun.
\citetalias{Brott10_gridpaper} argues that the surface gravity at the end
of the main sequence evolution for these objects is $\logg \simeq 3.2$ \citep[see also][]{Hunter08_vrot}.
While there are some observed stars with lower $\logg$, their nature is
unclear \citep{Vink10}.  We therefore exclude post-MS objects, by rejecting all
simulated stars with gravities below 3.2\,dex  (red
regime in Fig.~\ref{fig:selectioneffects}).  Note that this implies
that we also reject the final part of the main sequence evolution of stars
with masses above $\sim20$\Msun,
since the applied overshooting leads to an extension of the main sequence
to lower gravities for these objects.
We chose to do this because we favor a well defined selection
criterion and because the most massive objects have only a
low impact on our statistics. Most of the observed stars rejected are
luminosity class I supergiants, but this group also contains three
class II and two class III objects.

\subsubsection{Exclusion of Be stars}
\citet{Hunter09_nitrogen} did not derive nitrogen abundances for most
of the observed Be-type stars (the purple group in
Fig.~\ref{fig:mass-distribution}).  
For consistency, we excluded a fraction of the most rapid rotators in our
simulation. As the mechanism that produces the Be-phenomenon is still being discussed, 
it is difficult to define a clear criterion 
to identify simulated stars as Be stars. 
The Be mechanism  appears to be linked with 
fast rotation and pulsations \citep[e.g.][]{Cranmer05}. 
Commonly quoted required fractions of critical rotation for the occurrence of the  
Be phenomenon are 70-80\% \citep{Porter03}, 
though values as low as $\geq$~40\% \citep{Cranmer05} are also derived. 
\citet{Martayan08} finds $\Omega/\Omega_c$\,=\,85\% in the LMC, 
and \citet{Townsend04} argue that Be-stars might actually rotate as fast 
as 95\% of the breakup velocity, since the line broadening due to rotation
becomes insensitive to gravity darkening at high rotational velocities. 

The fraction of Be/(B+Be) in the observed sample is 13.5\%.
We can define a theoretical Be fraction within the limits of 
the yellow region of Fig.~\ref{fig:selectioneffects} by
dividing the number of stars rotating faster than a specified fraction
of break-up rotation by the number of all stars in that region.
To match this fraction with the observed one in our sample would require
a velocity limit of 55\% of the break-up velocity. 
 This number seems too low. As clearly more effects than fast rotation are involved 
in producing Be stars, we decided to not try to reproduce their number in the population synthesis.
In an attempt to simulate the effect of the exclusion of Be-stars from the sample,  
we chose a more conservative limit and excluded stars with a rotational velocity 
higher than 90\% of their breakup velocity in our simulations. 
This corresponds to  1.6\% of all stars in the yellow region of Fig.~\ref{fig:selectioneffects}.

In summary, we are unable to reproduce the fraction of observed Be-stars using our cut-off of
90\% of breakup velocity. However as we only exclude a small minority of stars from the
simulation, we do not expect that this will be a significant source of error.

\subsection{Stellar evolution models}
\label{sec:stev}

\begin{figure}[htbp]
\begin{center}
\includegraphics[angle=-90, width=0.5\textwidth]{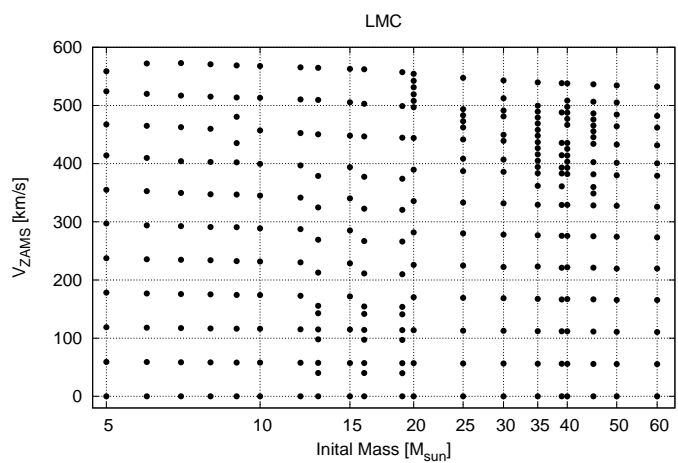}
\caption{Parameter space covered by the LMC  evolution grid. Each model sequence is represented by a dot
marking its initial mass and rotational velocity. }
\label{fig:grid}
\end{center}
\end{figure}

We calculated a grid of 284 stellar evolution models past core
hydrogen exhaustion using 
a one-dimensional hydrodynamic stellar evolution code, 
that includes the physics of differential rotation and magnetic fields
as described in \citetalias{Brott10_gridpaper}; see also
\citet{Yoon06} and \citet{deMink09}. The grid spans initial masses of
5--60\Msun and rotational velocities ($\vzams$) of 0--600\,$\kms$, as shown in
Fig.~\ref{fig:grid}. 
The grid is somewhat irregular in rotational velocity 
since we initialize our models as rigid rotators in a chemically homogeneous state
of thermal equilibrium. The subsequent adjustments to achieve CN-equilibrium in the core
alter the stellar structure on a thermal time scale and lead to a somewhat different surface rotation rate,
which thereafter changes only on the nuclear time scale, and which we designate as $\vzams$.

The effect of the centrifugal force on the stellar structure is considered
following \citet{EndalSofia76}. 
We consider the transport of angular momentum and chemical elements due to
various rotationally induced hydrodynamic instabilities, 
which include the Eddington-Sweet circulation, the dynamical and secular shear
instability, and the Goldreich-Schubert-Fricke instability \citep{Heger00}.
The transport of angular momentum by the Spruit-Tayler dynamo \citep{Spruit02}
is also implemented. 
Angular momentum transport and chemical mixing is approximated as a diffusive
process. 
 
Convection is treated by the
mixing length theory assuming a mixing-length parameter of
$\alpha_{\rm{MLT}}$\,=\,1.5.  We follow \citet{Langer91} for semi-convection,
for which  rather fast mixing is assumed, with a semi-convection efficiency
parameter of $\alpha_{\rm{SEM}}$=1.  
We consider convective overshooting using an overshooting parameter of $0.335\,\rm{H}_{\rm{P}}$. 
This value results from our
new calibration using the observed  $\vsini$ drop that is found in our data when
we plot $\vsini$ against the surface gravity \citepalias[see][]{Brott10_gridpaper} for more details).  
The rotational mixing efficiency has been calibrated such that the trends in
nitrogen of our sample are reproduced well. This leads to adopting
$f_{\rm{c}}$\,=\,0.0228, which is the ratio of the turbulent viscosity
to the diffusion coefficient~\citep[see][]{Heger00}.  

We use the recipe of \citet{Vink00, Vink01} for mass loss via stellar winds
from O- and B-type stars.  
To account for the effects of helium enrichment on the stellar wind,
we follow the approach of \citet{Yoon06} (see \citetalias{Brott10_gridpaper} for details).

\subsubsection{Initial chemical composition}

We choose an LMC mixture for the initial chemical composition of our
models (Table~\ref{tab:chem}). CNO abundances are taken from results for H\,{\sc  ii} 
regions \citep{Kurt98}. 
Mg and Si measurements for B-type stars from
\citet{Hunter08_vrot} and B-supergiants of \citet{Trundle07} are, on
average, 0.37\,dex lower than the solar values of \citet{Asplund05}.
Therefore, we have scaled all other elements from \citet{Asplund05} by
-0.4\,dex and adopted the average values of Mg and Si from
\citet{Hunter07_chem_Bstars} and \citet{Trundle07} for our stellar
models. The resulting metallicity of this mixture is $Z=0.0047$. We
have adopted the OPAL opacity tables \citep{opal}, for which we use
the Fe abundance to interpolate between tables of different
metallicities.

We have adopted the helium mass fraction using the primordial helium abundance
($Y_{\rm{p}}$\,$=$\,0.2477) from \citet{Peimbert07}.
Assuming the helium abundance to be a linear function of the metallicity $Z$
and fixing it to 0.28 at solar metallicity \citep{Grevesse96}
results in a value of 0.2562 for our LMC models.

 \begin{table}[tbp]
  \caption{LMC mixture used in our evolutionary models. }
  \centering
  \begin{tabular}{ l l |l l | l l}
     \hline\hline	
     C & 7.75 &   Mg    & 7.05              &  X &  0.7391\\
     N & 6.90 &    Si    & 7.20              &  Y & 0.2562\\
     O &  8.35 & Fe     &  7.05             &  Z  & 0.0047  \\
  \hline
\end{tabular}
\tablefoot{The baseline abundances of C, N, O,  Mg, Si and Fe are given in the first two columns, i.e. $\log(\rm{C/H})+12$. 
The total hydrogen (X), helium(Y) and metals (Z) mass fractions are given in the third column.}
\label{tab:chem}
\end{table}

More details on the calibration and choice of the chemical composition for the full
evolutionary model grid, which also includes models for Galactic and 
SMC stars, are presented in \citetalias{Brott10_gridpaper}. \\

\subsection{Effects of rotational mixing}
\label{sec:rotmix}

\begin{figure}[htbp]
\centering
\includegraphics[angle=-90,width=0.5\textwidth]{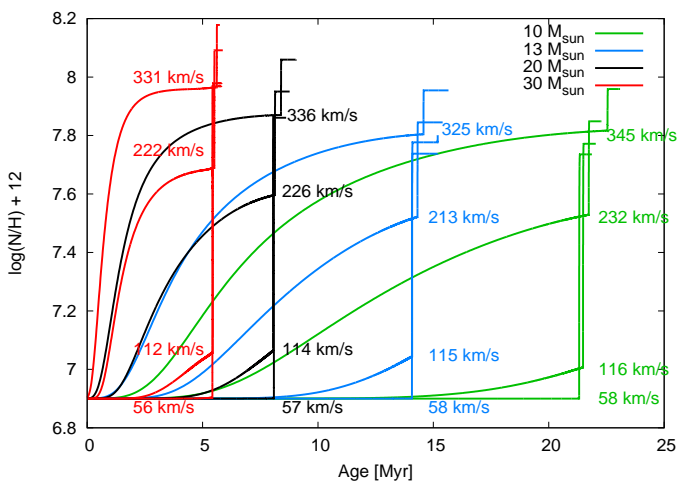}
\caption{\label{fig:nitrogen_time}Nitrogen surface abundance as a function of
time for models with different rotational velocities for 10\Msun (green),
13\Msun (blue), 20\Msun (black)  and 40\Msun (red) are shown. 
The model calculations have been ended shortly after the end of core hydrogen burning.  
The ZAMS velocities are indicated next to each track.  } 
\end{figure}

In our models, the strong transport of angular momentum due to the
Spruit-Tayler dynamo keeps the star rotating almost rigidly on the main
sequence.  As a consequence, the shear instability -- which is 
important for non-magnetic models~\citep{Heger00, Maeder00} --  
is less efficient than the Eddington-Sweet circulation in our models. 
\\
The surface abundances of stars can change due to rotational mixing on the main
sequence but not all elements are equally well suited for tracing such
mixing. While helium is produced in large amounts in the core, the models for
moderate rotational velocity show only small enhancements of helium at the
surface.  This is because the chemical stratification built up by hydrogen
burning between the helium enriched core and the hydrogen-rich envelope
effectively hinders chemical mixing across the boundary.  Only the fastest
rotating models ($\vzams\,>$ 350\,\kms) become significantly helium enriched at
the surface, and some of them even follow the quasi-chemically homogeneous
evolution if rotational mixing  operates faster than hydrogen burning
\citep{Yoon06}. Additionally, helium is already abundant at the stellar surface 
initially, and the expected enhancement relative to the initial
helium abundance is typically small.

Nitrogen, however, can be considerably enhanced at the  stellar
surface on the MS, even in relatively slow rotators. At the same time, 
carbon is depleted, but with a smaller relative change than nitrogen,
and oxygen remains almost constant.  This can be understood in the following
way. At the onset of the CNO-cycle, there is a short phase of the CN-cycle
in which most of the carbon is transformed to nitrogen.  
During this phase, hardly any helium is produced, and
the mean molecular weight gradient at the outer edge of the core
remains small. Therefore, nitrogen produced by the
CN-cycle can be efficiently transported into the layers above the convective
core. Once the mean molecular weight gradient
increases due to helium production, such mixing becomes much slower. 
As mixing throughout the radiative envelope --- i.e. above the
major mean molecular weight barrier produced by hydrogen burning ---
continues during the main sequence evolution,
the surface becomes gradually enriched with nitrogen. 
As the nitrogen abundance at the surface of early B~type stars can
be measured from optical spectroscopy, nitrogen is one of
the best tracers for rotationally-induced chemical mixing. 

In Fig.~\ref{fig:nitrogen_time} we show
the surface nitrogen abundance as a function of time
for a selection of models from our grid.
As the nitrogen enrichment during the main sequence evolution is a
monotonic function of time, nitrogen might be used as an age
indicator for stars with known initial stellar parameters. In our models, the
nitrogen abundance achieved at the end of core hydrogen burning 
is higher for higher initial rotation velocities, as the Eddington-Sweet
circulation operates at a higher flow velocity for faster rotation \citep{Yoon06}.
At the same time, at fixed initial rotation velocity, the maximum nitrogen abundance increases
only slightly with initial mass in the range of 10 to 20\Msun,
but it becomes more sensitive at masses above 30\Msun.
When hydrogen burning stops, the core
contracts and the envelope expands and eventually becomes 
convective, leading to a further dredge-up of nitrogen
and a sharp increase of the nitrogen surface abundance as shown in
Fig.~\ref{fig:nitrogen_time}. 

The abundances of other elements also show variations
due to rotational mixing. Sodium is enhanced due to proton captures in 
the NaNe-cycle, but it is difficult to determine observationally as its lines
are weak in this mass range and they are contaminated by interstellar absorption.
Boron is strongly depleted, because it is destroyed as it is mixed from the
relatively cool outer layers into hotter inner layers
\citepalias[see][]{Brott10_gridpaper}, but its spectral lines are
weak due to the low initial abundances. 
Carbon depletion can also be observed in the optical, but it 
cannot be measured as accurately as nitrogen due to NLTE effects
\citep{Hunter08_vrot}. Fluorine and aluminum are also affected.
A full account of this is given in \citetalias{Brott10_gridpaper}; in what follows we concentrate on 
the nitrogen abundances.

%: ------------------------- RESULTS ------------------------------------------------
\section{Results}
\label{sec:results}

\begin{figure*}[tp]
   \centering
	\includegraphics[angle=-90,width=0.8\textwidth]{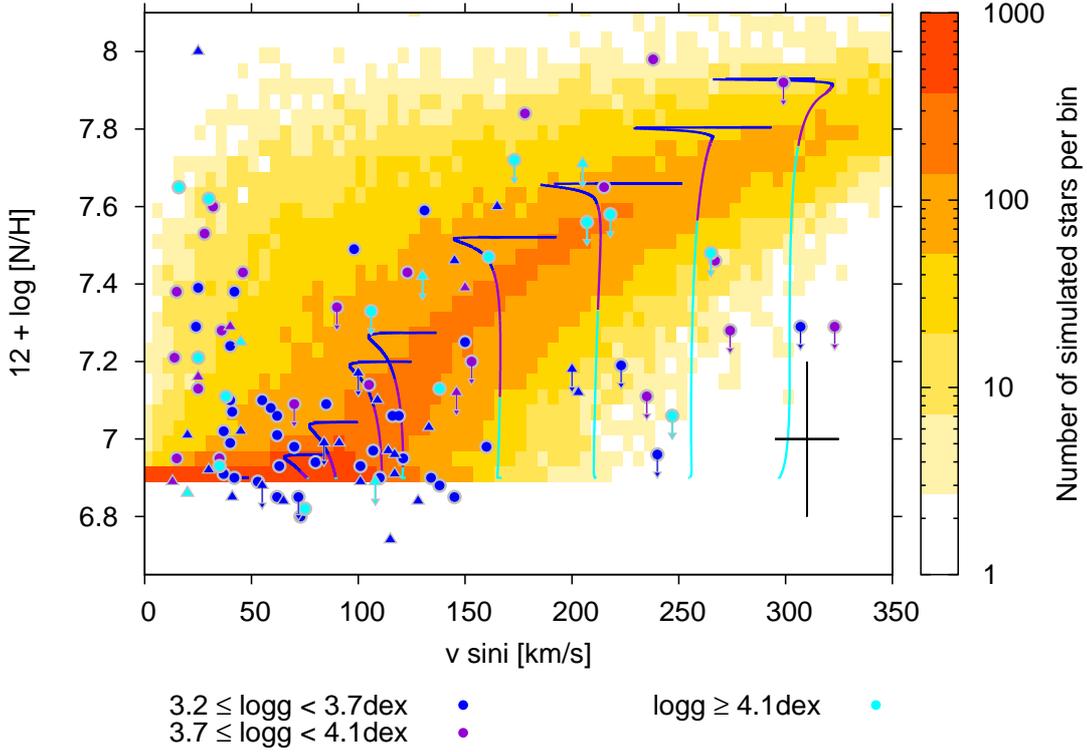}
	\caption{\label{fig:hunter_no_error} Hunter-plot showing projected rotational 
velocity against nitrogen enhancement. Our population synthesis is shown as 
a density plot in the background. The color coding corresponds to the number of stars per pixel. 
Overplotted are data from \citet{Hunter09_nitrogen}, with surface gravities 
as indicated by the color (see figure key). Single stars are plotted as circles, 
radial velocity variables as triangles. Evolutionary tracks of 13\Msun, corresponding to the average mass of
the sample stars, are shown 
with their surface gravity coded by the same colors as the observations. 
The velocities of the tracks have been multiplied by $\pi/4$, to account for the average projection effect.
The cross in the lower right corner shows the typical error on the observations.}
 \end{figure*}

\begin{figure*}[tp]
   \centering
	\includegraphics[angle=-90,width=0.8\textwidth]{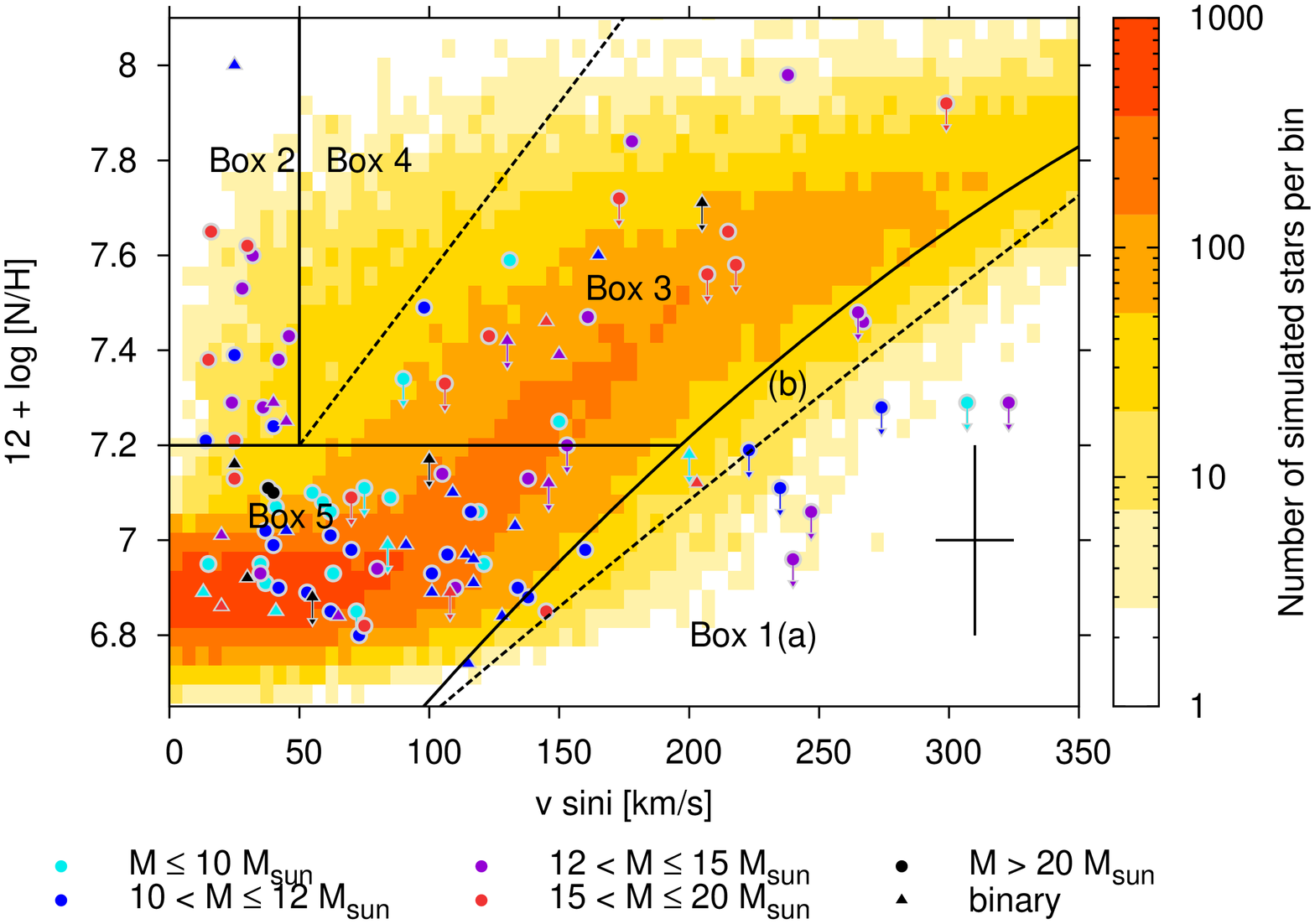}	
   \caption{\label{fig:hunter_with_magncutoff}
  As for Fig.~\ref{fig:hunter_no_error} but with a random error ($\sigma$ = 0.2\,dex) added to the nitrogen abundances 
in the simulated data. 
  The observational data have been color coded according to their evolutionary masses (see figure key). 
 For the boxes indicated, in Table~\ref{tab:boxes} we give the ratios of the number of observed to simulated stars} 
 \end{figure*}

To investigate the role of rotational mixing in massive stars, Fig.~\ref{fig:hunter_no_error} 
shows projected equatorial rotational velocities versus nitrogen abundance (hereafter the Hunter plot). 
The figure contains both observations and theoretical predictions. The observational 
data includes all stars from our LMC sample for which nitrogen abundances are available 
(see Sec.~\ref{sec:obs_sample}). 
Radial velocity variables are marked by triangles, while stars with 
no detected variations are represented with circles. 
The color coding of the symbol indicates the surface gravity of each object as shown.

The simulation considered 1.5 million stars with initial masses between 7 and
60\Msun. Only those that remained in the main sequence can be compared
to the observational sample. These are plotted in Fig.~\ref{fig:hunter_no_error}.
About 45.7\% of initial sample have left the main sequence, whilst another 46.3\% were excluded due
to selection criteria based on magnitude, temperature (spectral type) and surface gravity  
(see Sec.~\ref{sec:selectioneffects}). The remaining sample contains 120\,728 stars. 

The most important rejection criterion is the magnitude cutoff, but the age
of the simulated star also has a considerable effect. This is easily
understood considering that the maximum age of 35\,Myr in our SFH corresponds to the MS-lifetime of an
8\Msun model. 
As the MS-lifetime decreases rapidly with increasing initial mass
(see Fig.~\ref{fig:nitrogen_time}), the chance that a simulated star 
is assigned an age greater than its MS-lifetime becomes significant.

The magnitude cutoff eliminates a large number of simulated stars because
the IMF favors lower masses. For instance, models with {\em less} than 20\Msun 
are too faint to be observed for most of their MS-lifetime. A 12\Msun star spends 72\% of its 
MS-lifetime below the magnitude limit, while a 15\Msun star is not observable for 35\% of its MS-lifetime.
 
The simulated data have been grouped into $5\,\kms \, \times\, 0.04 \,\rm{dex}$ bins 
to generate a density plot. The color coding represents the number of stars in the bin 
(see the color bar on the right in Fig.~\ref{fig:hunter_no_error}). 
We also show evolutionary tracks of 13\,\Msun for comparison. For these tracks 
the surface gravity is indicated with the same color coding as for the observational data. 

Fig.~\ref{fig:hunter_with_magncutoff} shows the same population synthesis simulation as Fig.~\ref{fig:hunter_no_error}, 
but we have added a random Gaussian error with $\sigma = 0.2\,\rm{dex}$ to the predicted nitrogen
abundances. This error is characteristic of the spread in the nitrogen 
abundance determinations of the unenriched stars. Note that it
is smaller than the mean nitrogen error of the entire sample 
\citep[0.3--0.4 dex;][]{Hunter07_chem_Bstars,Hunter08_vrot}, which is represented 
by the cross in the lower right corner of Fig.~\ref{fig:hunter_no_error} and \ref{fig:hunter_with_magncutoff}.
We did not include an error for the predicted $\vsini$ values as the
observational values should be reasonably secure.
The color coding of the observational data in Fig.~\ref{fig:hunter_with_magncutoff} represents 
their estimated evolutionary mass. 

We have divided the diagram into five regimes:
\begin{description}
\item {\it Box 1(ab)} is a region of the diagram at high $\vsini$
  which is predicted to be almost empty. The full line defining box 1(ab)
  gives the locus where the number of simulated stars per pixel is
  $\sim 40$, while this number is $\sim 15$ for the dashed line
  defining Box~1b.

\item {\it Box 2} is a region of the diagram where very few stars are predicted, but where a group of enhanced, 
slowly rotating stars are observed. This group of stars was already described by \citet{Hunter09_nitrogen},
and will be discussed further below.

\item {\it Box 3} encloses the stars in our simulation which have a clear
signal of nitrogen enhancement due to rotational mixing.

\item {\it Box 4 } is a region where neither stars are observed, nor a significant number of them is predicted.

\item {\it Box 5} is the region of the diagram where the un-enriched stars are located.
The initial nitrogen abundance is about 6.9 with a typical spread of about 0.2\,dex.
We chose the upper boundary of 7.2 such that the probability of observed stars found
above this line to be truly enriched is large.
 
\end{description}
\begin{table}
	\centering
	\caption{
	Fraction (in per cent) and absolute number of observed and simulated stars found in each box defined in 
Fig.~\ref{fig:hunter_with_magncutoff}. }
		\begin{tabular}{l|rr|rr|rr}
		\hline\hline
		& \multicolumn{4}{c|}{observed}  & \multicolumn{2}{c}{modeled} \\
		& \multicolumn{2}{c}{all} & \multicolumn{2}{c|}{excl. UL} & \multicolumn{2}{c}{} \\
		&\multicolumn{1}{c}{\%} & \multicolumn{1}{c|}{\#} & \multicolumn{1}{c}{\%} & 
                \multicolumn{1}{c|}{\#} & \multicolumn{1}{c}{\%}& \multicolumn{1}{c}{\#$ / 10^{3}$} \\
		\hline
		Box 1ab & 14.0 & 15 &  7.4 &  6 & $4.6^{+0.8}_{-1.0}$  &  $5.6$\\[1.5pt]
		Box 1a  &  5.6 &  9 &  0   &  0 & $1.6^{+0.5}_{-0.5}$  &  $2.0$\\[1.5pt]
		Box 1b  &  8.4 &  6 &  7.4 &  6 & $3.0 ^{+0.3}_{-0.5}$ &  $3.6$\\[1.5pt]
		Box 2   & 15.0 & 16 & 19.8 & 16 & $1.2^{+0.0}_{-0.1}$  &  $1.4$\\[1.5pt]
		Box 3   & 17.8 & 19 & 13.6 & 11 & $35.7^{+4.0}_{-4.6}$ & $43.1$\\[1.5pt]
		Box 4   &  --- &    &  --- &    & $2.6^{+0.3}_{-0.5}$  &  $3.1$\\[1.5pt]
		Box 5   & 53.3 & 57 & 59.3 & 48 & $55.9^{-5.2}_{+6.2}$ & $67.5$\\[1.5pt]
	       \hline
                Total   &      &107 &      & 81 &                      &$120.7$\\[1.5pt]       
		\hline
		\end{tabular}
		\tablefoot{For the observed stars, in Column~2 we take all stars into account, 
in Column~3 we only consider those without upper limits. 
The third column gives the expected number of stars from our population synthesis model. 
Errors on the number of simulated (Column~6) stars are derived by comparing to simulations with a velocity 
distribution that was broader or narrower by $\pm25\,\kms$. }
	\label{tab:boxes}
\end{table}

In Table~\ref{tab:boxes} we present the absolute and relative number of stars populating each box 
in both the observations and simulations. For the observations we present statistics 
including and excluding objects for which only an upper limit to the nitrogen abundance was established.
A downward revision of the upper limits might reduce the number of observed stars in Box~3,
while that in Boxes~5 and~1 might increase. 
For all stars in Box\,1a we only have upper limits, 
but their presence in this box is not affected by this. 
Unless otherwise stated, the numbers quoted in the discussion below include the stars
with upper limits to their nitrogen abundances.

We find 53.3\% of the total observed sample to be in Box\,5,
containing the unenriched stars which do not rotate rapidly. 
This is in good agreement with the theoretical prediction of 55.9\%.
Due to the large number of stars in this region, this agreement is
not affected by stars with upper limits on their nitrogen abundance. 
Most of the stars in this box are expected to have a relatively low mass,
and in an advanced phase of core hydrogen burning
such that they have become sufficiently bright in the optical to be above the magnitude cutoff.
Indeed the observed stars with  $\rm{M}\leq$ 10\Msun accumulate in Box\,5
and appear to have mainly surface gravities of less than 3.7.

We note that a more efficient mixing would
reduce the predicted number of stars in Box~5. However, the mixing efficiency is calibrated
to best represent the correlation between nitrogen abundance and $\vsini$ (see above).
{\em The fact that after such a calibration the number of non enriched stars predicted
in Box\,5 is in very good agreement with the observations supports  the theory of rotational mixing.}

\begin{figure}[htbp] 
   \centering
   \includegraphics[angle=-90,width=0.5\textwidth]{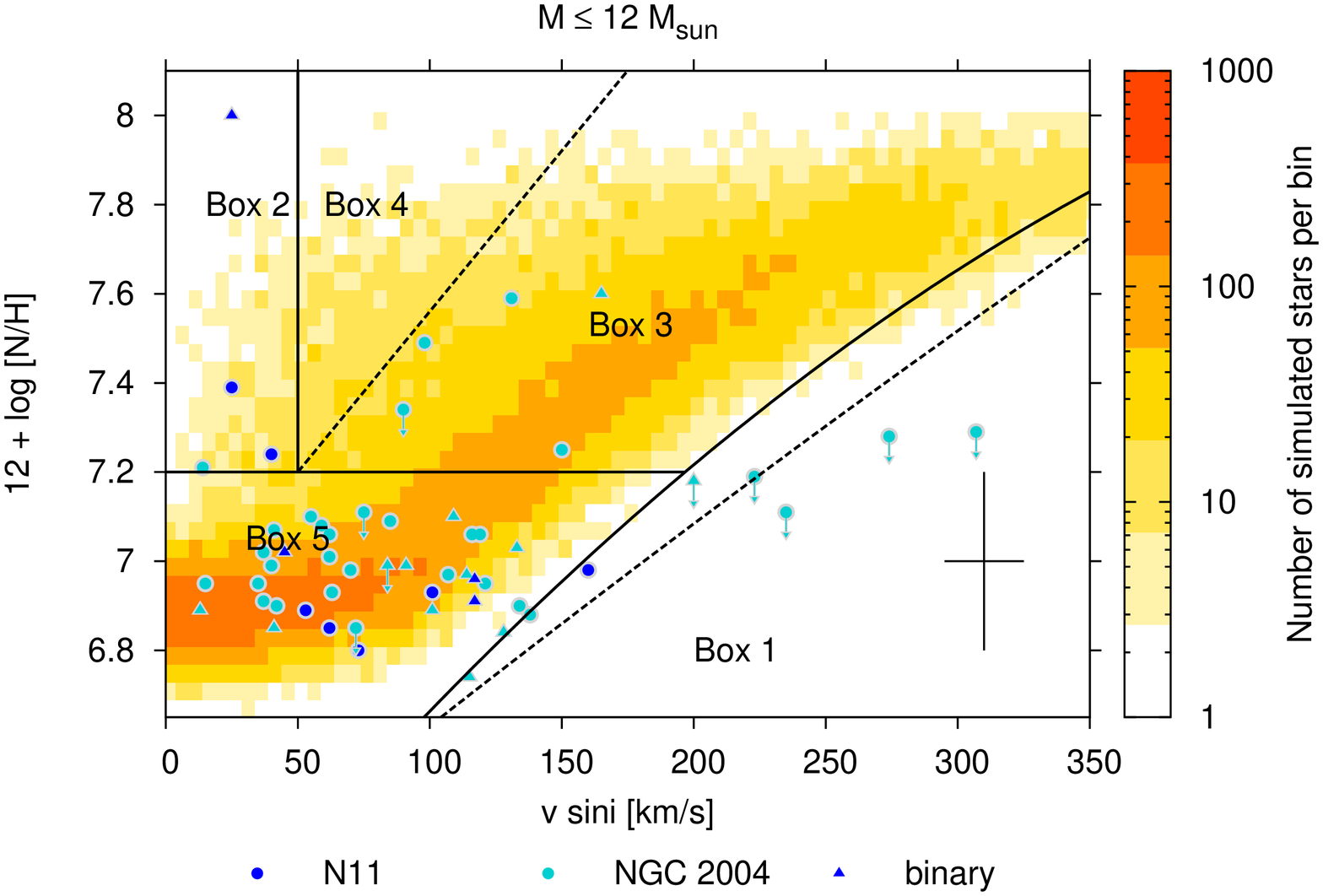} 
   \includegraphics[angle=-90,width=0.5\textwidth]{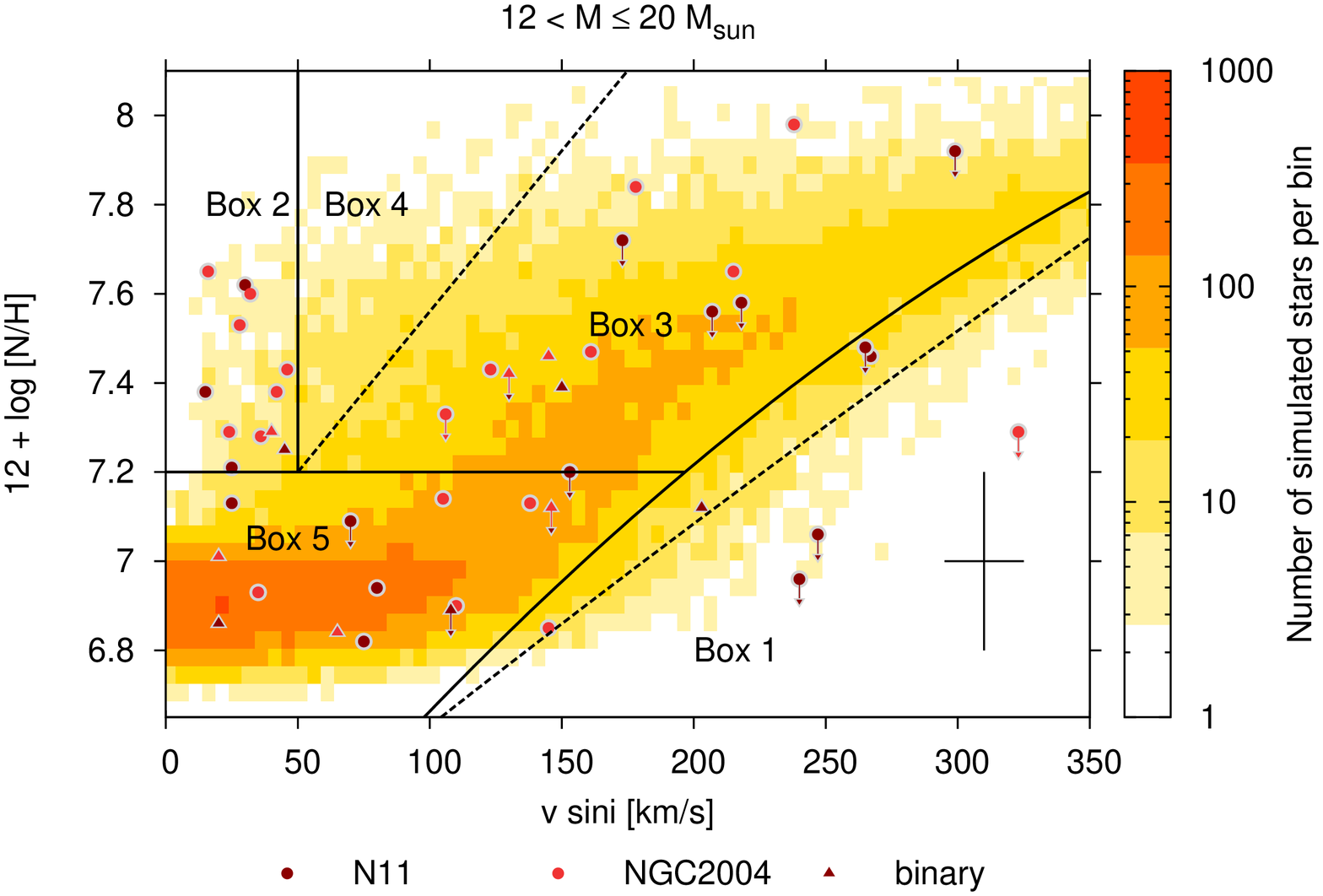} 
   \includegraphics[angle=-90,width=0.5\textwidth]{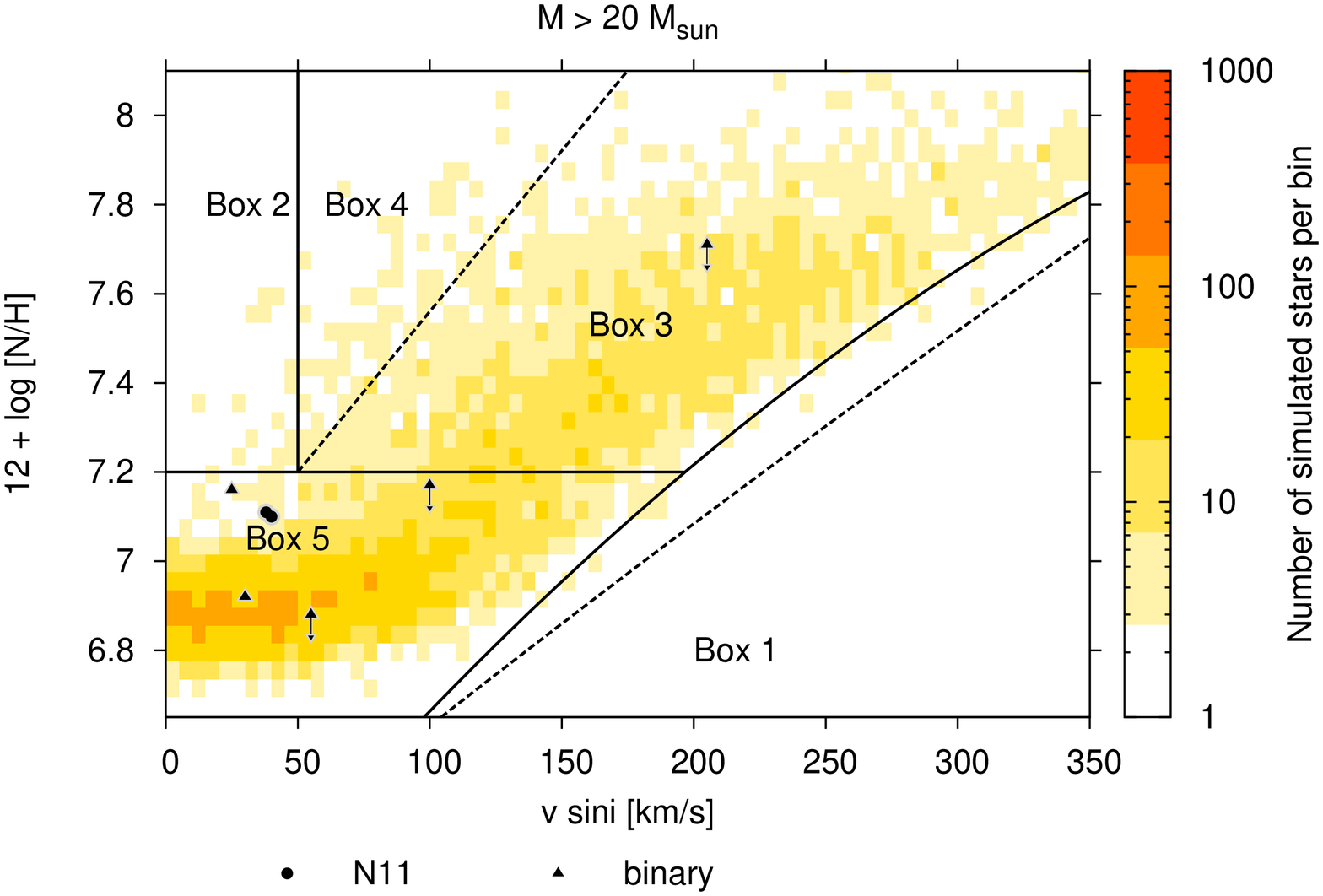} 
   \caption{\label{fig:hunter_massbins}
Mass dependence of the Hunter-plot. We have separated 
Fig.~\ref{fig:hunter_with_magncutoff} into three mass bins. Top to bottom: 
Population synthesis results and observational data for the mass bins 
$\rm{M}\leq 12$\Msun, $12\leq \rm{M} <20$\Msun, $\rm{M} >20$\Msun are shown, respectively. 
   }
\end{figure}

Box~3, containing the stars of our simulation which show rotationally induced nitrogen enhancement,
is predicted to contain 35.7\% of our sample. However
only 17.8\% of the observed sample is found in this region. 
Hence we predict approximately twice as many stars to be enriched as are actually observed.
Subtracting the stars for which we have only upper limits leaves only
13.6\% of the sample stars in Box~3, increasing the discrepancy between theory
and observations to almost a factor of three. 
On the other hand, Fig.~\ref{fig:hunter_massbins} shows that more observed stars in Box~3 are 
found in the 12...20$\mso$ bin compared to the lower mass bin ($M < 12\mso$),
while our models predict  similar (absolute) numbers for both mass bins.
This may indicate that subdividing our sample into too small subsamples
results in a loss of statistical significance.

While this may show that analyzing subsamples may lead to effects
of small number statistics, it also shows that the overall discrepancy
in the predicted and observed number of stars in Box~3 may not be very significant. 

Box 2 contains a group of apparently slowly rotating but nitrogen enhanced stars. 
Theoretical models of rotating single stars predict such enhancements 
only for fast intrinsic rotation ($\vzams > 150\kms$).
Note that all the observed targets in Box 2
cannot be rapid rotators seen nearly pole on. Indeed our simulations (which assume random
inclination angles) predict that only 1.2\% of the sample (corresponding to one star) would
be found in this region. Observationally 16 stars comprising 15\% of the sample lie in Box 2.
Therefore the projection effect cannot populate this part of the diagram.
This statement is enforced by the fact that no stars are observed in Box~4,
which --- if the projection effect were to explain the large number of stars
in Box~2 --- should contain more stars than found in Box~2. 
We conclude that the vast majority of stars in Box~2 are intrinsically
slow rotators and their high nitrogen abundance is not explained by
our models. 
The nitrogen enhanced stars in Box\,2  appear to have higher masses and to be less evolved.
There are no stars with  $\rm{M}\leq$ 10\Msun in this group, even though this is one of the most populated
mass bins in the observed sample. With the exception of five stars (of which one is a binary)
surface gravities in this group are larger than 3.7\,dex. A comparison with the evolutionary tracks in
Fig.~\ref{fig:hunter_no_error} indicates that most stars of this box
are in an intermediate or evolved stage of their MS-evolution.
{\em The large number of stars found in Box\,2
is not predicted by the theory of rotational mixing in single stars
with an overpopulation of observed targets by more than a factor of ten with that predicted.}

Box~1(a+b) contains a group of 15 fast rotating but unenriched stars (14\% of the observed sample). 
We predict a factor of 3 less stars (4.6\%) in this box, with most of them
expected to be in Box~1b, i.e. very close to Boxes~3 and~5.
In particular the area where the six observed stars in Box~1a are found is 
predicted to be empty.  This prediction is largely due to the adopted observational
selection effects, but to a small extent also due to the fast nitrogen enrichment of rapidly rotating stars
(Fig.~\ref{fig:nitrogen_time}).
Most of the simulated targets populating the diagram started their
evolution below the magnitude cutoff. 
Those which were rapidly rotating have sufficient
time (before passing the magnitude limit) to enhance their surface nitrogen
abundance so significantly that once they appear in the diagram they show up in Box\,3. 
In Box\,1(a or a+b), there does not seem to be a preference for high surface gravities and low masses
as suggested by \citet{Maeder09_coast}. The six objects in Box~1a have 
masses in the range of $9\ldots15$\Msun, which is a
typical spread for the sample. Their average mass is 12\Msun which is similar
to the average for all the stars for which nitrogen abundances have been determined of 13.6\Msun.
Additionally the stars cover a range of ages.
As the stars in Box 1b have upper limits to
their nitrogen abundances, some of them could actually lie in Box 1a, increasing the
discrepancy with prediction.
{\em In summary the existence of a significant number of targets in
Box~1a cannot be reconciled with the prediction of rotational mixing for single stars that
this area should contain effectively no objects.}

Fig.~\ref{fig:hunter_massbins} indicates that whilst the population of Box~1b is strongly mass dependent,
that dependency is less pronounced for Box~1a.
The top panel shows simulation and data with $\rm{M}\leq 12$\Msun.
The models of 12\Msun
spend about 70\% of their MS at luminosities below the magnitude limit.
When they finally become bright enough, they are already so enriched that they show up in Box~3.
Only for models of intermediate mass, as shown in the middle panel (12--20\Msun),
does the interplay between the selection effects and the mixing-timescale allow stars to populate Box~1.
Using an even finer mass binning, we find that Box~1a is actually only populated
by stars in the mass range 12--15\Msun, while Box 1b contains models in the mass range 12--20\Msun.
The bottom panel shows models and observed stars more massive than 20\Msun.
In both cases Box~1 remains empty. Models of 20\Msun or more are above the magnitude limit all time,
but they have effective temperatures above 35$\,$000\,K during their early main sequence evolution,
such that the fast rotators, once they have become cool enough to appear in our simulation data,
have already achieved a significant nitrogen enrichment.

\subsection{Model uncertainties}
\label{sec:uncertainties}
Before we focus on the possible evolutionary implication of our findings, 
we first discuss the significance of the results summarized in Table~\ref{tab:boxes}
which are obtained from our population synthesis. 
The discrepancy between observation and prediction is very large for both Boxes~1 and~2. In
contrast, it is smaller for Box~3, being only a factor of approximately two. Hence in
discussing our uncertainties, we will focus on how they might affect our simulation of the
fractional population of this Box.

These numbers are subject to various uncertainties: 
{\it  i)} the mixing efficiency in the stellar models,
{\it ii)} the initial velocity distribution, mass function and star formation history,
{\it iii)} the observational biases, and
{\it iv)} exclusion of unsuitable stars.

\subsubsection{Mixing efficiency}

The calibration of the mixing efficiency is described in detail in \citetalias{Brott10_gridpaper}.
The models in our stellar evolution grid are calculated with a mixing efficiency parameter of
$f_{c}=2.28\times10^{-2}$ (see Sec.~\ref{sec:stev}). The impact of a larger mixing parameter $f_{c}$ is
that a larger maximum nitrogen enhancement is achieved for a given initial rotation rate.
Therefore, the general trend of nitrogen enhancement versus $\vsini$ would be shifted to 
lower rotation velocities and larger nitrogen abundances.  
In Fig.~\ref{fig:hunter_no_error}, 
the evolutionary tracks of our 13\Msun models show 
an unaffected nitrogen abundance for $\vsini < 100\, \kms$
within the observational error,
and a peak value of 7.9 is obtained at $\vsini \simeq 300\,\kms$.
This behavior does reflect the general trend of the observed stars
in Boxes~5 and~3 quite well: the maximum observed nitrogen abundance,
the slope of the nitrogen versus $\vsini$-relation, and the
foot-point at $\vsini \simeq 100\, \kms$ are all matched well.
We thus believe that the calibration of the mixing efficiency of our models
does not contribute significantly to the observed discrepancy.

\subsubsection{Input distribution functions}

The adopted distribution of initial rotational velocities has 
been discussed in Sec.~\ref{sec:vinit}. In order to assess the sensitivity of the predictions on 
the distribution of initial rotational velocities, we investigated the effect of 
a broader ($\sigma = 166\,\kms$) and narrower ($\sigma = 116\,\kms $) width of the underlying Gaussian. 
A variation of $\pm25\,\kms$ is consistent with the errors given in \citet{Dufton06_vrot} 
on the fit to the projected rotational velocity distribution. 
The relative changes are given as errors on the theoretical percentages in Table~\ref{tab:boxes}.
While we find that the numbers predicted for most boxes do not change significantly,
the narrower velocity distribution leading to fewer rapid rotators, 
shifts about 6\% of all modeled stars from Box~3 to Box~5.
This reduces the discrepancy for Box~3 (35.7\% predicted vs. 17.8\% observed)
to below a factor of two (29.8\% predicted vs. 17.8\% observed),
but it cannot eliminate it.

The initial mass function is assumed to be Salpeter ($\alpha=-2.35$; see Sec.~\ref{sec:imf}). 
Given the apparent universal nature of this distribution in the mass range 
from 10--60\Msun\citep{Chabrier03,Kroupa01},
and given the good agreement between the observed 
and simulated mass functions (see Fig.~\ref{fig:IMF}) 
we believe that this is unlikely to be a significant source of error.

As discussed in Sec.~\ref{sec:sfh}, we have assumed continuous star formation
in our population synthesis model. 
Clearly, the star formation history may influence the distribution of 
stars in Fig.~\ref{fig:hunter_with_magncutoff}. 
For instance, a young co-coeval population will cause the fraction of predicted stars in Box~5 to increase,
at the expense of the number of stars in Box~3.
However, as Fig.~\ref{fig:isochrones} appears largely consistent with 
our assumption of continuous star formation, we do not expect a significant error
due to this assumption.  
We have investigated the case where star formation stopped after 25\,Myr 
(as suggested in Sec~\ref{sec:initial_distributions}) instead of continuing for 35\,Myr.
The stars that are older than 25\,Myr are mainly to be found in Boxes\,3 and\,5. 
If we exclude these stars, the number of stars in Boxes\,3 and\,5 change by about 1\%, 
and by much less in the other boxes.  

\subsubsection{Observational biases}

While we accounted for the observational biases in our population 
synthesis model, some error may be associated with this. Most relevant in this
respect may be the magnitude cut-off applied in the VLT-FLAMES Survey.
While Fig.~\ref{fig:selectioneffects} shows that indeed no observed stars
fainter than our adopted magnitude cut-off are found in the sample,
there may be a slight deficit of observed stars near the magnitude cut-off,
especially at masses below $\sim 10\,$\Msun. As these stars would be nearly 
at the end of their main sequence evolution, the ratio of enriched to unenriched
stars amongst them would be greater than average. 
However, limiting the population synthesis to stars above 9\,\Msun has a very similar 
effect as limiting the star formation history to 25\,Myrs (discussed above),
since the main-sequence lifetime of a non-rotating 9\Msun model is about 25\,Myrs.
It changes the number of stars in Box\,3 and~5 only by less than one percent. 
The main-sequence lifetime of a non-rotating 9\Msun model is about 25\,Myrs. 

Binary stars are slightly
brighter than single stars of a mass or brightness equal to the binary primary. Therefore,
some binaries with primaries which would not pass the magnitude limit by themselves
might, through the additional light of their companion.
A test population simulation where we assumed all stars to consist of binary stars
with identical companions convinced us that this effect does not significantly affect
any results described below.

For very rapid rotators, the observed brightness and gravity may also depend
on the inclination angle due to the latitudinal dependence of gravity darkening.
However, the observed and the modeled number of very rapid rotators
is so small that this effect is negligible here \citep[see also][]{Hunter09_nitrogen}.

\subsubsection{Removal of unsuitable stars}

In our population synthesis simulation, we have only included single stars computed
with a single set of physics assumptions. In reality however, different physical situations
may occur. For example, many massive stars are known to have a close binary companion
\citep[e.g.][]{Sana09}
which will modify the evolution during core hydrogen burning.
Other massive stars are clearly affected by internal magnetic fields
\citep[e.g.][]{Wade06}. It remains unclear wether the magnetic fields used in our evolutionary models for angular momentum transport describe those detected in massive stars.
Our  population synthesis model is therefore only representative of
at most a certain fraction of the observed stars. 

In order to assess the potential consequences for our conclusions concerning
the number of stars in Box~3, let us first assume that the stars in Boxes~1
and~2, which cannot be explained well by our model anyway, need to be
discarded as ``unsuitable'' for a comparison with our model.
This would reduce the total number of stars to be compared with
from~107 to~76, and then render the fraction of observed stars in 
Box~3 to~25\% (rather than 17.8\%), which is closer to the predicted
fraction of 35.7\%. 
 
However, the fraction of observed stars
which is ``unsuitable'' for a comparison with our model may be greater;
let us assume it is as high as 50\%.
As the observed stars in Box~3 agree well with the physics of rotational mixing,
let us further assume that none of the 50 or so observed stars to be disregarded
would lie in Box~3.  Again, we would assume that the observed stars in Boxes~1 and~2
should be disregarded. However, about 20 stars need to be disregarded also
from Box~5. For example, this could be fast rotators which cannot mix because
of strong internal magnetic fields. With such assumptions, we had 37 ``suitable'' stars
left in Box~5, and 19 in Box~3. The latter number corresponds to $\sim$33\%
of the reduced sample, which then
is close to the predicted fraction of 35.7\%. While the assumption that we can
describe only half of the stars in the observed sample may be extreme, 
this case cannot be ruled out, and it serves as an example 
of the difficulty in defining a robust observational sample.

\subsubsection{Stochastic effects}
Here, we assess the spread in the number of stars obtained for each box in Fig.~\ref{fig:hunter_with_magncutoff} 
if only 107 stars are drawn randomly from the simulated population. 
This is equivalent to the following experiment. A pot contains N=120\,728 balls, equivalent to the number of 
simulated stars in Fig.~\ref{fig:hunter_with_magncutoff}. The balls in the pot have the colors c$_{1}\dots$c$_{5}$. 
The number of balls in the pot with color c$_{i}$ is equal to the number of simulated stars in Box\,i, 
for $i=1\ldots 5$.
We draw n=107 balls from this pot, without taking the order into account in which they are drawn.
Because of the large number of balls in the pot, the probability $p_{i}$ to pick a certain 
color can be approximated as constant, i.e. as the number of balls of color c$_{i}$ divided by the total number of balls N. This approximation is valid as long the number of balls of color c$_{i}$ is much larger than the number of balls drawn from the pot. 

When the experiment is repeated many times, the result is multinominally distributed\citep[][p. 75]{Papoulis84}.
The expectation values $\mu_i$ for the number balls drawn from each color and its standard deviation $\sigma_i$ can be calculated by: 
\begin{align}
\mu_{i}&= np_{i}& \label{eq:mu}\\
\sigma^{2}_{i} &= np_{i}(1-p_{i}) \label{eq:sigma}
\end{align}
The expectation values given in Table~\ref{tab:expectation-values} agree 
with the percentages found in Table~\ref{tab:boxes}. The standard deviation indicates a rather small spread, 
which cannot cause the discrepancy between observation and simulation discussed above. This is true in particular for the ratio of the number of stars from Box\,3 and Box\,5. 

\begin{table}
\caption{Expectation value of the numbers of stars and standard deviation for each box} 
\label{tab:expectation-values}
\centering
\begin{tabular}{l|c|c}
\hline\hline
& $\mu \pm \sigma$[\#]  & $\mu \pm \sigma$[\%] \\
\hline
Box\,1 & 4.9 $\pm$ 2.2 & 4.6 $\pm$ 2.0 \\
Box\,2 & 1.3 $\pm$ 1.1 & 1.2 $\pm$ 1.0 \\
Box\,3 & 38.2 $\pm$ 5.0 & 35.7 $\pm$ 4.6 \\
Box\,4 & 2.8 $\pm$ 1.6 & 2.6 $\pm$ 1.5 \\
Box\,5 & 59.8 $\pm$ 5.1 & 55.9 $\pm$ 4.8 \\
\end{tabular}
\tablefoot{Expectation values and standard deviations as calculated from equations~\ref{eq:mu} and \ref{eq:sigma} are given in column one. In the second column these values are expressed in percent of the sample.  }
\end{table}

\subsubsection{Overall assessment}

We have shown above that the discrepancy of a factor of two in the predicted and observed fraction
of stars in Box~3 is not sufficient to rule out rotational mixing for this subsample.
Our theoretical mixing calibration, the adopted star formation 
history, and initial distribution functions for mass and rotational velocity apparently
induce little uncertainty. However, the adopted magnitude cut-off and in particular
the possibility of applying different physical situations to main sequence stars than
assumed in our model allows for a modification of the fractions of stars
in Boxes~3 and ~5 which may bring theory and observations into agreement. 

On the other hand, the possible agreement
of the predicted distribution of stars in Boxes~3 and~5 is not sufficient to
verify the theory of rotational mixing. This will be discussed further when
predictions of massive close binary evolution are included in our considerations,
in the next section.

%-------------------------- DISCUSSION ------------------------------------------- 

\section{Discussion}
\label{sec:discussion}

From the previous section, we conclude that overall, 
the agreement between the population synthesis model and the observed sample
is rather poor. While the model predicts stars to be confined to Boxes~5 and~3, the observed stars
are much more wide spread. 
Quantitatively, only in Box~5 is the number of observed stars in good agreement with the predictions.
While the stars in Box~3 at least qualitatively agree with
the population synthesis model, the stars in Box~1 and Box~2 are completely unexpected and 
cannot be explained by the model. Before any further conclusion, it is thus worthwhile
to consider physical and evolutionary processes which could populate Boxes~1 and~2, as those might
have effects on Boxes~3 and~4.

\subsection{Close binary evolution}
\label{sec:binarys}
Though our model assumption is that the diagram is populated only by single stars,
it is well known that a significant fraction of stars  is part of a binary system 
\citep[e.g.,][]{Mason09}, with a binary fraction of 0.5 or more 
found in young star clusters \citep[e.g.,][]{Sana08, Sana09, Bosch09}, and with a 
period distribution such that many of them will interact \citep{Sana10}.
We emphasize that strong binary interaction (i.e., mass transfer 
or merging) will often lead to a system that is either a single star
or which shows insignificant radial velocity variations.
Most observed radial velocity
variables in the VLT-FLAMES Survey will likely be pre-interaction systems
and thus well suited for comparison with single star evolution.
The apparent binary fraction in our various observational subsamples thus give
no direct clue to the importance of strong binary interaction for the subsample. 

In close binaries, the primary star may fill its Roche lobe during the main sequence evolution. 
This results in the removal of the stellar envelope of the primary star, 
which is transformed into a (mostly unobservable) helium star. 
If the entire envelope is added to the companion, the surface nitrogen abundance of this star 
will increase by a factor of~3 to~5, even when rotational mixing 
is not taken into account \citep[cf., Fig.~9 in][]{Langer08_iaus} as in the accretion process the envelope
material is well mixed. 
Subsequent rotational mixing of the mass gainer may enhance
the surface nitrogen abundance further on the thermonuclear
timescale \citep[cf., Fig.~10 in][]{Langer08_iaus}. 
If only a small fraction of the material is accreted, the result is no or only a mild N enrichment. 
During the accretion process not only mass but also angular momentum is transferred to the companion, 
and it has been found that the accretion of only about 10\% of 
the stellar mass is able to spin-up the star significantly \citep{Packet81, Langer08_iaus}. 

Therefore, conservative and moderately non-conservative mass transfer 
would tend to populate the upper right corner 
in Fig.~\ref{fig:hunter_with_magncutoff} (Box\,3), 
while highly non-conservative mass transfer 
\citep[e.g.,][]{Petrovic05} is likely to 
move stars to the lower right corner (Box\,1) \citep{Langer08_iaus}.  
Binary interaction leading to a merger is also expected to result in fast 
rotating objects, likely populating Box\,1 or Box\.3.
Binary evolution is also able to populate Box~2, as post-mass transfer systems may remain so tight
that the tidal interaction can spin down the accretion star \citep{Langer08_iaus}.

While quantitative statements requires binary population 
synthesis models which include a methodology for non-conservative mass transfer 
(which are not yet available), it is clear that the canonical mass transfer 
evolution leads to a rapidly rotating, nitrogen-rich main sequence accretion star
\citep{Langer08_iaus}.  
Thus, invoking binaries has the potential to solve the problems
with Boxes~1 and~2, but for each star appearing in these boxes, one or more stars would appear
in Box~3 due to binary interaction, as binary evolution is expected to lead more often into Box~3 
than into Box~1 or~2.

This implies that, using binaries to solve the problems
with Box~1a and, in particular, with Box~2, worsens the problem with Box~3, possibly even to the
extent that no rotationally mixed single stars are required to explain those found in Box~3. 
{\em We conclude that close binary evolution, as far as it is currently understood, 
is unlikely to be responsible for the slowly rotating nitrogen-rich population
of observed stars in Box~2.}

\subsection{Magnetic fields}

Effects of magnetic fields other than the magnetic angular momentum transport 
implemented in our models may affect a fraction of massive stars.
Further down the main sequence,
magnetic Ap stars tend to rotate more slowly than non-magnetic A stars and show surface abundance anomalies. 
Also the ratio of Ap to A stars is about 5-10\% \citep{Wolff68} which is not far 
from the percentage of stars which populates Box\,2. Furthermore, \citet{Morel08} recently
found magnetic fields in Galactic nitrogen enriched slowly-rotating early B-type dwarfs. 
All this makes the possibility that magnetic fields play a role in the population of stars in Box\,2  
an interesting direction of investigation.

A potential magnetic mechanism would not necessarily need 
to act simultaneously on both the nitrogen abundance and the stellar rotation,  
as is believed to occur in Ap stars. In fact one could envisage a mechanism that slows stars down
through magnetic braking, with rotational mixing occurring whilst the star is still a fast rotator 
responsible for the nitrogen enrichment. 
The fact that the number of predicted stars in Box\,3 is more or less the 
sum of the stars observed in Boxes\,3 and\,2 
may be suggestive of a migration process from Box\,3 to Box\,2. 
The lack of observed stars in Box~4 then indicates that such a migration 
needs to occur on a rapid time scale. 
A massive star example for magnetic braking has been found by \citet{Townsend10} in the 
magnetic helium-strong star $\sigma$ Ori E with a characteristic spin down time of 1.34\,Myr. 
Also, \citet{Meynet11} have calculated models with a simple description of magnetic braking
that leads to slow rotation and produces a nitrogen surface enhancement
by inducing strong internal differential rotation.

On the other hand, one may also imagine a migration of stars from Box~5 to Box~2,
e.g. through the buoyant rise of magnetic bubbles which carry magnetic flux and nitrogen from the 
convective core to the surface \citep{MacGregor03}. 
{\em We conclude that magnetic field effects may play a role, but
they are not sufficiently understood to allow secure predictions.}

\subsection{Comparison with earlier analyses}

\begin{figure}[htbp]
\begin{center}
\includegraphics[angle=-90,width=0.5\textwidth]{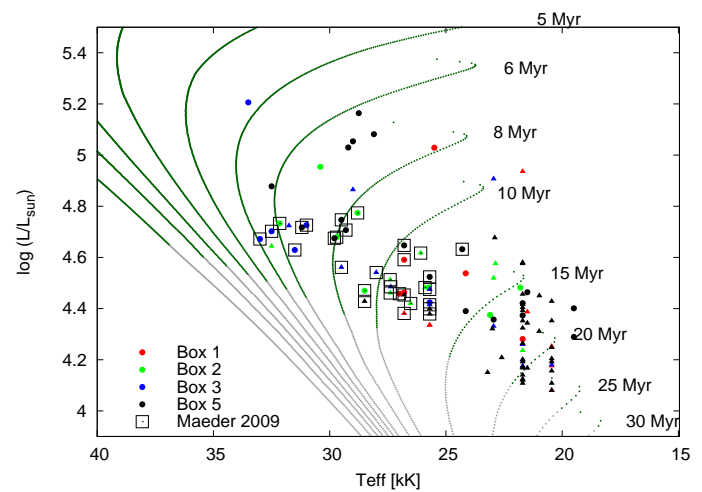}
\caption{HR-diagram of the 107 sample stars in the Hunter diagram. Stars from the N11 and NGC 2004 fields 
are represented by circles and triangles, respectively. The color coding represents the box into which stars fall in the Hunter diagram. Isochrones made from non-rotating evolution models are shown in green. The isochrone part below the magnitude cut is plotted in gray. }
\label{fig:MaederSelect}
\end{center}
\end{figure}

Some of the conclusions derived in this paper were already obtained by \citet{Hunter08_letter}, based on the
same observational sample as the one scrutinized here, but only based on a subset of the stellar
models presented in \citetalias{Brott10_gridpaper} rather than on detailed population synthesis.
\citet{Maeder09_coast} argued that the conclusions of Hunter et al. were unwarranted, due to
the dispersion of the sample stars in mass, age, rotation, and binarity
(we ignore the suggestion of a metallicity spread, which is known to be small in the LMC).
In their own analysis, where they reduced the sample size by a factor of about~4 to consider
only stars of similar mass, they
still maintained the dispersion in age (see Fig.~\ref{fig:MaederSelect}), rotation and binarity, 
and found their models of rotating single stars to explain the observations satisfactorily.

We find the picture emerging from the reduced sample analyzed by \citet{Maeder09_coast} and that
from \citet{Hunter08_letter} not very different. 
This is in fact not surprising, since the restriction in mass applied by Maeder et al.
is not necessary. The mass range of stars in the full sample is already so small
(9...25\Msun), that the expected nitrogen enrichment is very similar throughout 
the considered mass range (cf. Fig.~8, and also
Fig.~1 of \citet{Maeder09_coast}). This is in fact confirmed by their analysis,
which shows the bands of potentially rotationally mixed stars in 
Figs.~3 and~4 of \citet{Maeder09_coast} being exactly the same, implying that the 
mass dependence is negligible. While Maeder et al.
conclude that the stars in Box~3 agree with single star models, we do the same, only we note that
binary evolution provides an alternative explanation for these stars. 

Maeder et al. also
find stars in Box~2 in their reduced samples, with a similar proportion as they occur in the
full sample. While Maeder et al. argue that they may occur due to binary effects and thus
do not disturb the agreement of single star predictions with the observations, 
\citet{Meynet11} argue that massive slowly rotating nitrogen-rich main sequence stars could
be single stars undergoing magnetic braking, as proposed by \citet{Hunter08_letter}.
Only the fraction of rapidly rotating stars without strong nitrogen enrichment (Box~1)
in the \citet{Maeder09_coast} sample is smaller than in the full sample, which is due to
the small size of the samples of Maeder et al. However, of course, all Box~1 stars in the full sample
remain to be explained.

In any case, we developed the population synthesis approach as presented in this paper 
in order to construct a tool to analyze also diverse stellar samples in a statistically
sound way. The spread in age, mass and rotation (and in principle, metallicity) is fully 
accounted for by our method. Our new results reinforce the
uncomfortable conclusions of \citet{Hunter08_letter} that two significant groups of massive core hydrogen burning
stars stand out as being in conflict with evolutionary models, posing a challenge to
the concept of rotational mixing.

%: -------------------- CONCLUSIONS ------------------------------------------
\section{Conclusions}
\label{sec:conclusion}
Through detailed single star population synthesis modeling, we have tried to reproduce 
the properties of the
largest homogeneous sample of early B-type stars in the LMC.
This sample is the first for which nitrogen surface abundances have been
determined for a wide range in projected rotational velocities.
Uncertain mixing parameters in the underlying stellar evolution models
(overshooting and rotational mixing) have been calibrated such that they 
reproduce the observed main-sequence widening deduced from the 
sharp drop in rotational velocities (\citetalias{Brott10_gridpaper}, \citealp{Vink10}),
as well as the observed pattern of nitrogen enrichment as a function of
the projected rotation velocity in the majority of fast rotators \citep{Hunter08_letter}.

We confirm quantitatively the result of \citet{Hunter08_letter}
that two sub-populations of stars, namely rapidly rotating, unenriched
stars and slowly rotating, nitrogen-rich stars (Boxes~1a and~2 in Fig.~\ref{fig:hunter_with_magncutoff}),
are not reproduced by a population synthesis model for single stars.

While the group of rapidly rotating nitrogen-rich stars (Box~3) is, 
by construction, in qualitative agreement with
our simulation, only about half of the number of predicted stars are found
to be present in the observational sample. Although possibly explained
by uncertainties in our method (see Sec.~\ref{sec:uncertainties}),  
predictions of close binary evolution models, which also produce
nitrogen-rich rapid rotators,
tend to increase this discrepancy.

We find it unlikely that the group of 
slowly rotating nitrogen-rich stars (Box~2 in Fig.~\ref{fig:hunter_with_magncutoff})
is produced via mass transfer and subsequent tidal spin-down in close binary systems,
as a binary population which
could produce these stars would likely overpopulate the group of
rapidly rotating nitrogen-rich stars (see Sec.~\ref{sec:binarys}).  
At present, it appears plausible that some of the unexplained
slowly rotating, nitrogen-rich stars 
are affected by magnetic fields \citep[cf. the magnetic
field measurements for Galactic stars with similar properties;][]{Morel08}.

While a detailed physical prescription of possible magnetic effects
in stellar models is currently not available, several binary effects are
much better understood. Therefore, it appears promising to perform
a binary population synthesis of massive main sequence stars in the
near future, which will give quantitative constraints on the relative
importance of binaries in the various subgroups. In fact, this appears
to be required in order to arrive at solid conclusions on the nature
of the group of rapidly rotation nitrogen-rich stars, 
and, thereby, obtain clear evidence for the universality
of rotational mixing in massive stars.

Further light should be shed on this by the ongoing Tarantula
Survey \citep{Evans10_tarantula}, which has targeted over 1000 stars
in the 30\,Doradus of the LMC. 30\,Dor is the largest H\,{\sc ii} region in
the Local Group, containing a large number of very massive stars.  The
new survey is tailored for the detection of binarity, so will help
sample the mass range above 30\Msun, where mixing effects should be
more significants, and therefore more easily detectable.

\acknowledgements{
We thank Georges Meynet, the referee of this paper, for
many helpful comments and suggestions.
SdM  acknowledges support for this work provided by 
NASA through Hubble Fellowship grant
HST-HF-51270.01-A awarded by the Space Telescope Science Institute,
which is operated by the Association of Universities for Research in
Astronomy, Inc., for NASA, under contract NAS 5-26555.}

%:------------------- Literature ---------------------------------------

\bibliography{../../literatur}

%:------------------Appendix ----------------------------------------------------
\appendix
\section{The LMC Sample}
\label{App:A}
\begin{figure*}[htbp]
	\centering		
	\begin{tabular}{cc}
	\includegraphics[angle=-90,width=0.5\textwidth]{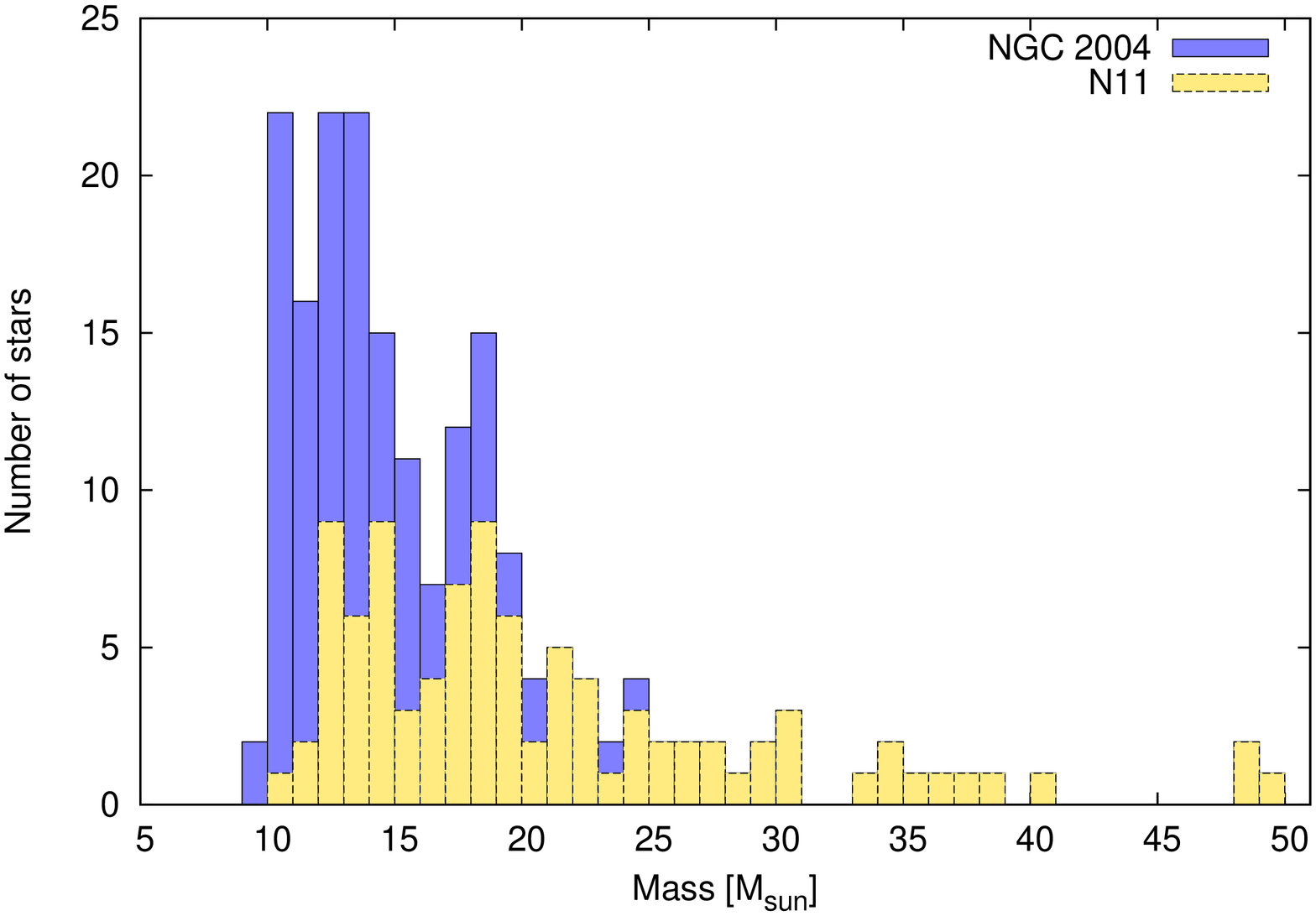} &
	\includegraphics[angle=-90,width=0.5\textwidth]{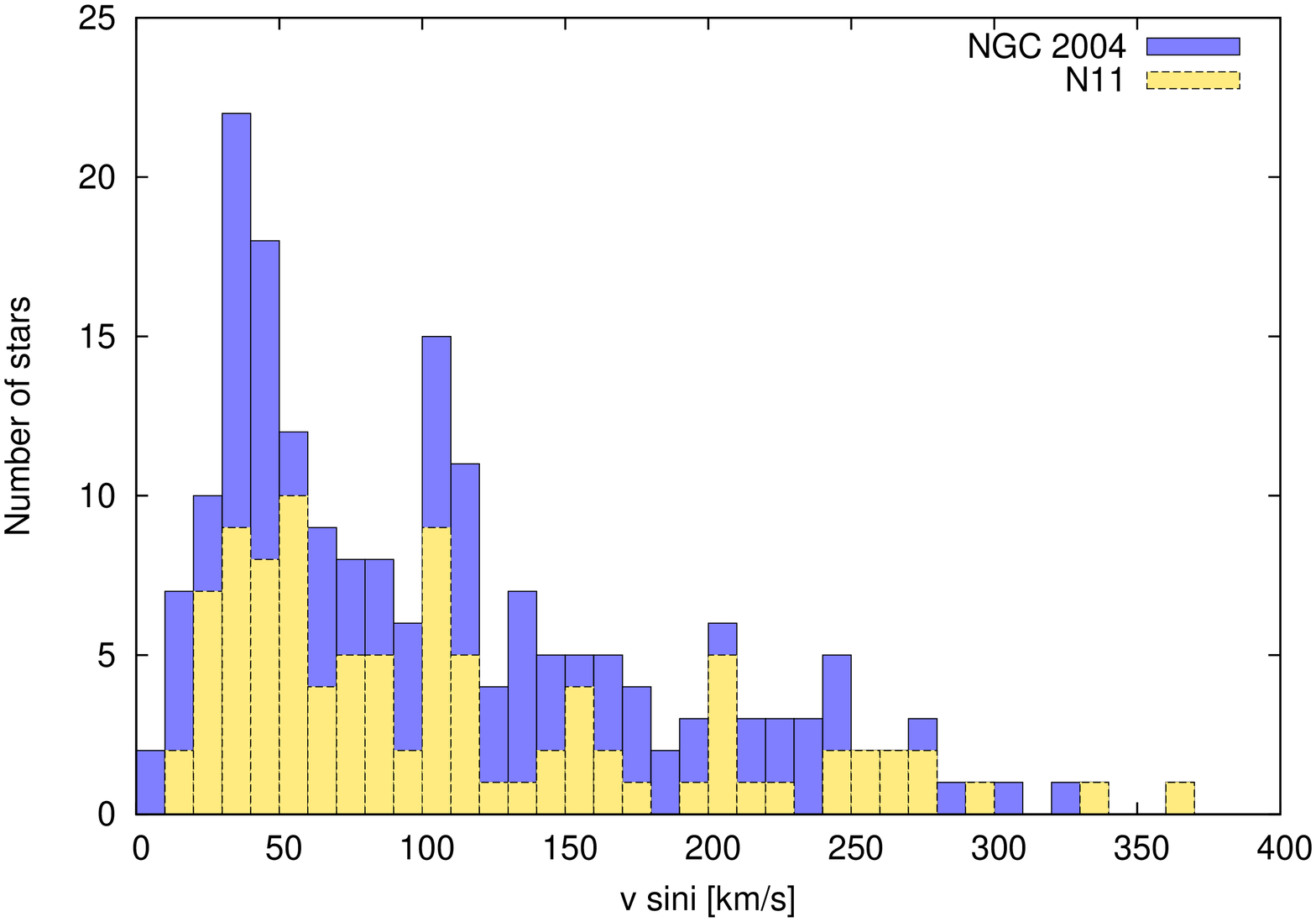}\\
	\includegraphics[angle=-90,width=0.5\textwidth]{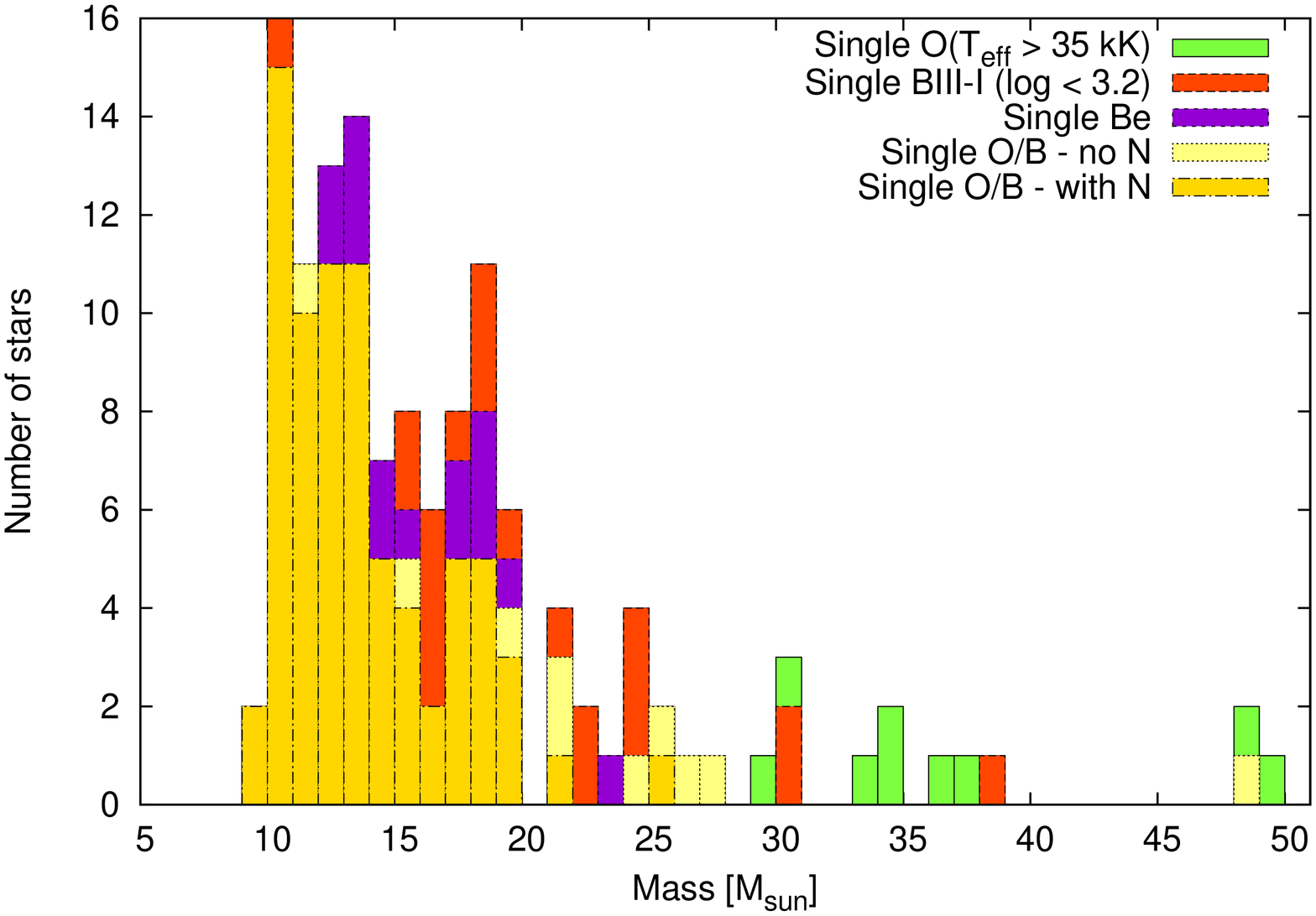}&
	  \includegraphics[angle=-90,width=0.5\textwidth]{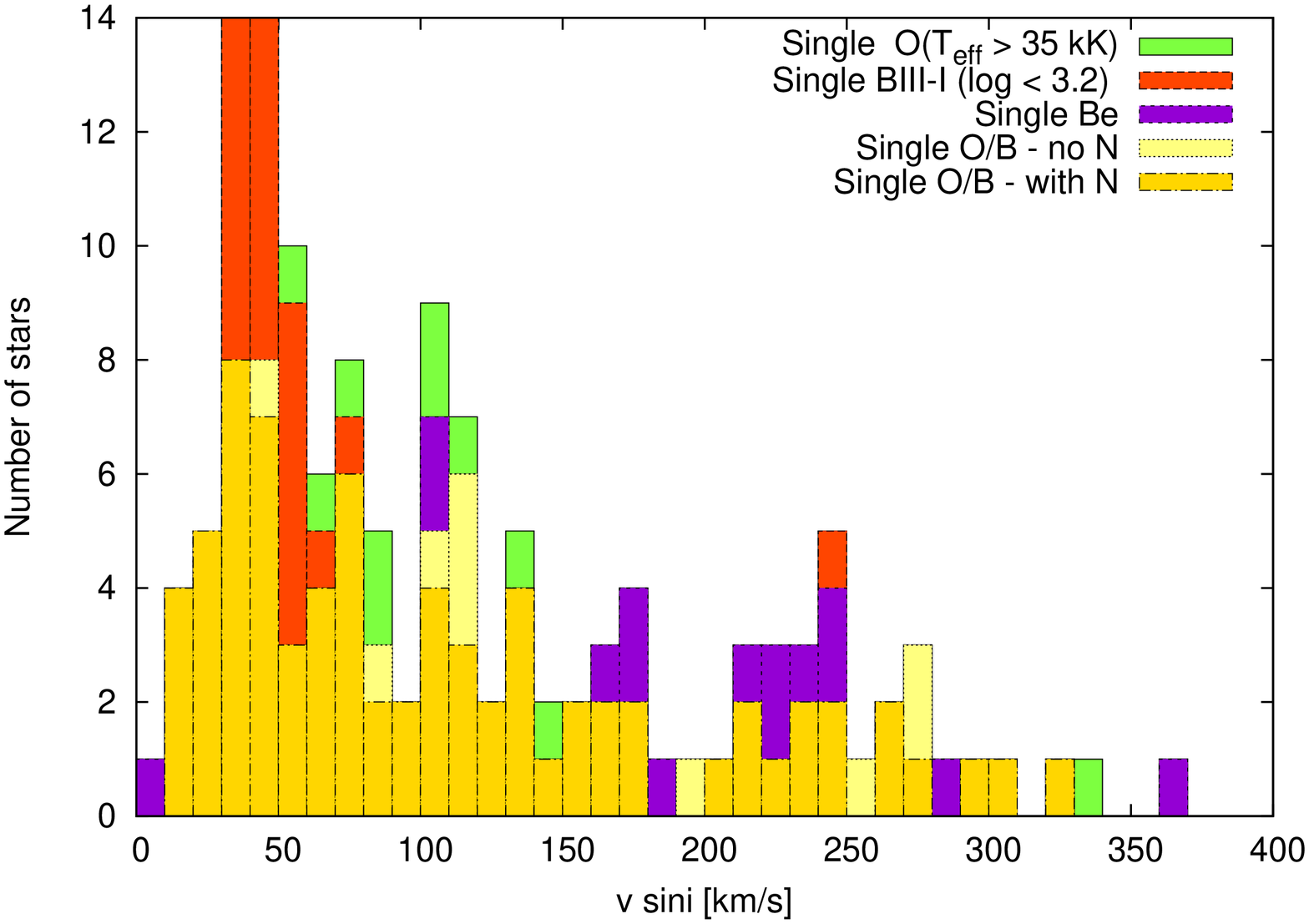} \\	
	\includegraphics[angle=-90,width=0.5\textwidth]{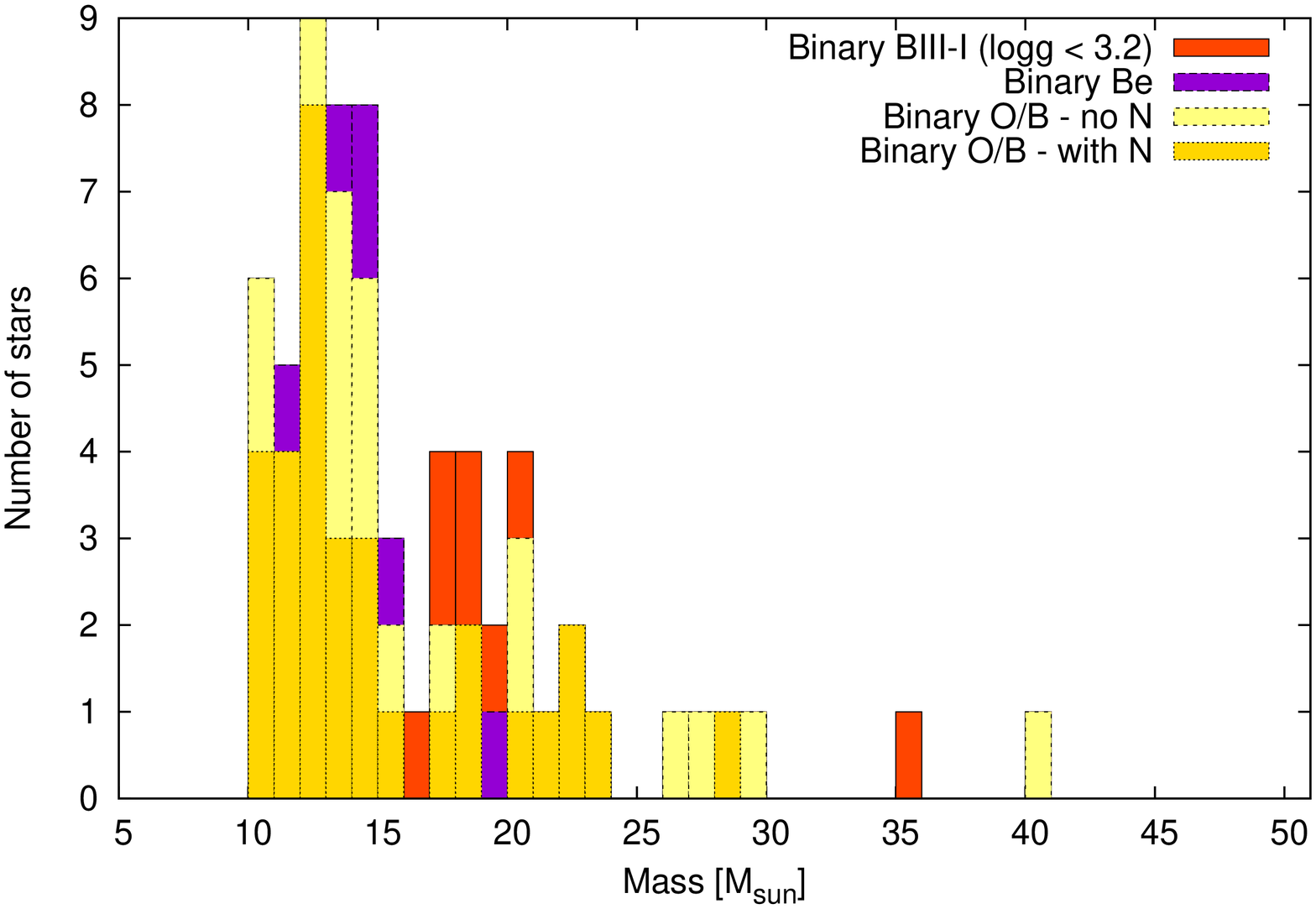}&	
	\includegraphics[angle=-90,width=0.5\textwidth]{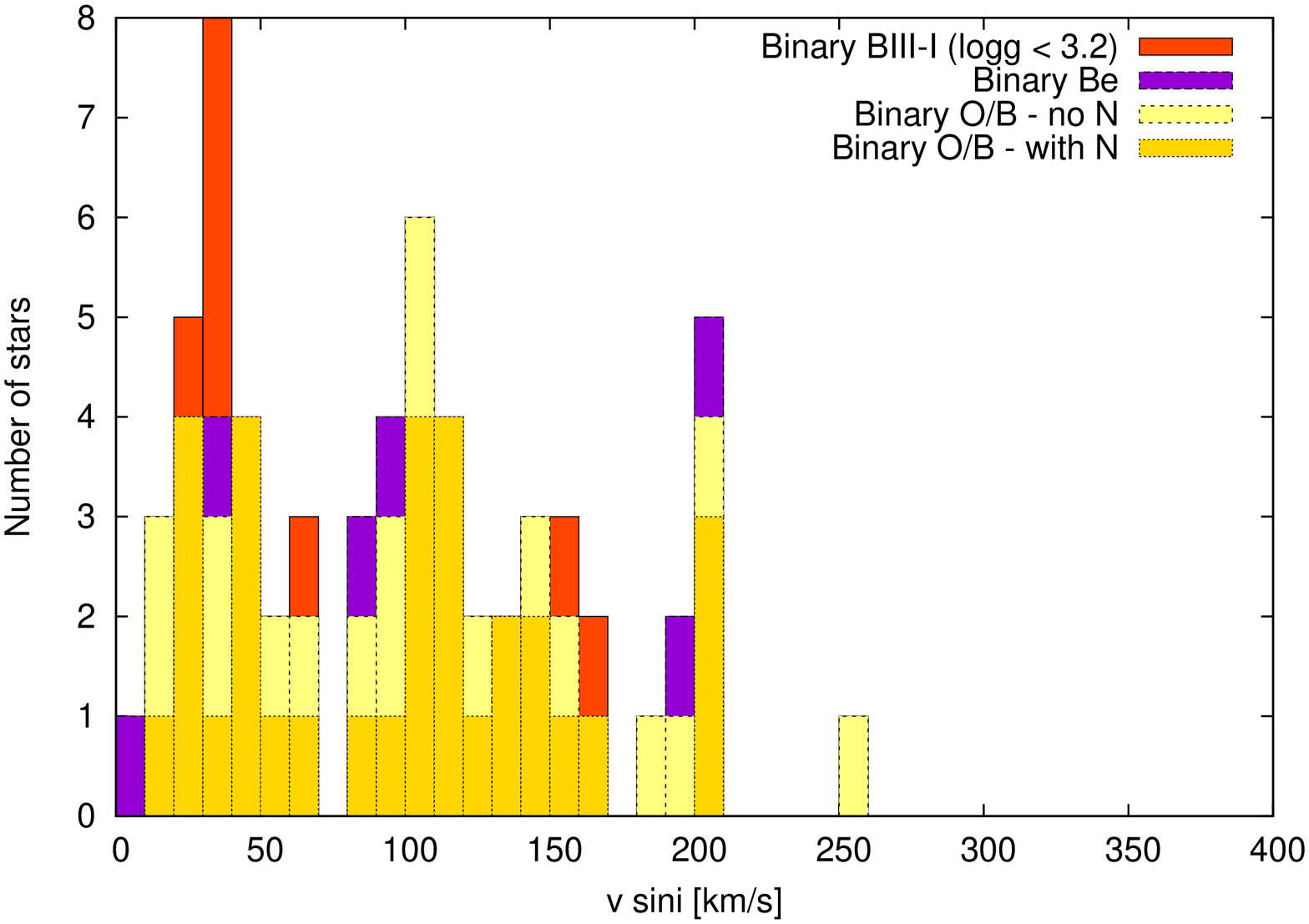} 	  
	\end{tabular}

\caption{Distribution of evolutionary masses (left) and projected rotational velocities (right) of O- and B-type stars from \citet{Hunter09_nitrogen} in the 
NGC\,2004 and N11 VLT-FLAMES fields.  {\it Top panels:} The contributions from each VLT-FLAMES field are shown in yellow (N11) and blue (NGC 2004); {\it Middle:} Apparently single stars; 
{\it Bottom:} Radial velocity variables.  The selection effects discussed in Sec~\ref{sec:selectioneffects} are color-coded as per the legend in each plot.
 For the objects in light yellow we do not have nitrogen measurements for various reasons. The stars shown with gold panels are those with measured nitrogen abundances
that are the primary focus of the current analysis.
Not shown in the mass histogram are two O-type stars, with masses of 62 and 83\,\Msun. } 
	\label{fig:vsini-distribution}
	\label{fig:mass-distribution}
\end{figure*}

In Fig.~\ref{fig:mass-distribution} we show the mass and projected rotational velocity distributions of our sample. 
 In the upper panels, the targets are divided into the VLT-FLAMES fields. The rotational velocity distributions in both fields are very similar. The total mass distribution is dominated by N11 stars at high masses, while NGC\,2004 contributes mainly to the lower-mass end. 
 
We divided our sample into single and binary stars, as shown in the middle and lower 
panels of Fig.~\ref{fig:mass-distribution}, respectively.  
The color coding corresponds to our selection effects that are later applied to each star 
of our population synthesis  (see discussion in Sec.~\ref{sec:selectioneffects}). 
The single star velocity distribution is well represented by those stars 
with measured nitrogen abundances (yellow panels). 
Around 50\,\kms is a peak of low gravity objects (orange), 
which is somewhat sensitive to our overshooting calibration. 
A value of 50\,\kms is also  comparable to the expected the line broadening due to macro turbulence, 
therefore, these stars could be rotating at lower velocities. 

The mass distribution of single stars with measured nitrogen abundances is also representative of the mass distribution of the total single star sample. Nitrogen abundances were not determined for the limited sample of the most massive stars owing to the problems of modeling the N\,{\sc iii} line.
The binary velocity distribution shows three peaks (at 30, 100 and 200\,\kms), 
which may well be not statistically significant given the small number of stars.

\end{document}